\documentclass[%
 reprint,
superscriptaddress,
citeautoscript,
nofootinbib,
nobibnotes,
 amsmath,amssymb,
 aps,
prb,
]{revtex4-1}
\setcitestyle{numbers,square}
\usepackage[protrusion=true,expansion=true]{microtype}
\usepackage{times}
\usepackage{graphicx,bm,xcolor,braket, tikz}
\usepackage[normalem]{ulem}
\usepackage{hyperref}
\hypersetup{
    pdftitle={main},    
    pdfauthor={Tarantelli-Scopa},
    colorlinks=true,
    citecolor=blue,
    linkcolor=blue,
    urlcolor=blue
  }
\usepackage{amssymb,physics}
\usepackage[utf8]{inputenc}
\usepackage[english]{babel}

\makeatletter
\def\graphicscale{\twocolumn@sw{0.3}{0.4}}
\def\graphicthreescale{\twocolumn@sw{0.3}{0.4}}

\newcommand{\be}{\begin{equation}}
\newcommand{\ee}{\end{equation}}
\newcommand{\bea}{\begin{align}}
\newcommand{\eea}{\end{align}}
\newcommand{\de}{\partial}

\usepackage{dsfont}
\usepackage{mathrsfs}
\begin{document}

\title{Out-of-equilibrium scaling behavior arising during round-trip protocols across a quantum first-order transition}

\author{Francesco Tarantelli}
\email{francesco.tarantelli@phd.unipi.it}
\affiliation{Dipartimento di Fisica dell’Università di Pisa and INFN, 
Largo Pontecorvo 3, I-56127 Pisa, Italy}

\author{Stefano Scopa}
\email{sscopa@sissa.it} 
\affiliation{SISSA and INFN, via Bonomea 265, 34136 Trieste, Italy}

\date{\today}

\begin{abstract}
We investigate the nonequilibrium dynamics of quantum spin chains during a round-trip protocol that slowly drives the system across a quantum first-order transition. Out-of-equilibrium scaling behaviors \`a la Kibble-Zurek for the single-passage protocol across the first-order transition have been previously determined. Here, we show that such scaling relations persist when the driving protocol is inverted and the transition is approached again by a far-from-equilibrium state. This results in a {\it quasi-universality} of the scaling functions, which keep some dependence on the details of the protocol at the inversion time. We explicitly determine such quasi-universal scaling functions by employing an effective two-level description of the many-body system near the transition. We discuss the validity of this approximation and how this relates to the observed scaling regime. Although our results apply to generic systems, we focus on the prototypical example of a $1D$ transverse field Ising model in the ferromagnetic regime, which we drive across the first-order transitions through a time-dependent longitudinal field.

\end{abstract}
\maketitle
\section{Introduction}
Quantum spin chains featuring two quasi-degenerate vacua in competition are recently attracting a great deal of attention in the context of lattice gauge theories. One of the goals of these studies is to shed light on some ununderstood high-energy physics phenomena --such as the false vacuum decay \cite{kuhn2015non,pichler2016real,buyens2017real,magnifico2020real,lagnese2021false,milsted2022collisions} or the confinement mechanisms \cite{martinez2016real,yang2020observation,simon2011quantum,vovrosh2021confinement}-- in a controllable way, thanks to the impressive development of modern quantum simulators. Out of equilibrium, these efforts provided, for instance, a characterization of the unusual spreading of correlations and entanglement \cite{kormos2017real,lerose2020quasilocalized,tortora2020relaxation,lagnese2022quenches,scopa2022entanglement,castro2020entanglement,vovrosh2021confinement,rigobello2021entanglement}, as well as of the thermalization \cite{birnkammer2022prethermalization,james2019nonthermal,robinson2019signatures,chanda2020confinement}, in condensed-matter analogs of confined systems. A celebrated example is the Ising model in a tilted magnetic field \cite{mccoy1978two,delfino1996non,fonseca2003ising,rutkevich2008energy,lake2010confinement,coldea2010quantum}, where topological excitations in the ferromagnetic phase are subject to an effective confining force induced by the longitudinal field, which make them behaving as toy version of mesonic excitations \cite{surace2021scattering,vovrosh2022confinement,karpov2022spatiotemporal,vovrosh2022dynamical}.\\

From a different perspective, it is well known that quasi-degenerate vacua naturally arise in the context of quantum phase transitions, after a spontaneous symmetry breaking. Their behavior and coexistence in the non-critical regime is governed by a first-order transition (FOT). 
FOTs are responsible for many important out-of-equilibrium effects, including nucleations and metastability \cite{binder1987theory,bray2002theory}, coarsening \cite{chandran2012kibble}, and anomalous dependence on the boundary conditions \cite{pelissetto2020scaling,panagopoulos2018dynamic,campostrini2015quantum,pelissetto2018finite,rossini2018ground}. Moreover, a nice analogous at first-order transitions of the Kibble-Zurek mechanism \cite{zurek2005dynamics,dziarmaga2005dynamics,polkovnikov2005universal,dziarmaga2010dynamics} (see also Refs.~\cite{kibble1976topology,kibble1980some,zurek1985cosmological,zurek1996cosmological} for the classical formulation) has been proposed by Vicari {\it et al.}~\cite{campostrini2014finite,pelissetto2017dynamic}, recently reviewed in Refs.~\cite{rossini2021coherent,pelissetto2023scaling}. The essential idea is that one can construct a finite-size scaling theory during the slow driving of the non-critical model across the transition upon replacing the diverging correlation length and time with the corresponding typical scales characterizing the FOT.\\
Under this analogy, as well as the Kibble-Zurek mechanism captures the defects density generated across the criticality from an initial equilibrium homogeneous state, out-of-equilibrium finite-size scaling relations at the FOT quantify the transition to the first excited level during the driving from an initial (non-critical) ground state. However, understanding whether similar scaling relations occur when the system is driven across a quantum phase transition from an out-of-equilibrium configuration is still not clear. So far, results are limited to the recent Ref.~\cite{tarantelli2022out}, and yet unexplored for FOTs. This is the scope of this paper. Below, we investigate the emergence of finite-size scaling behaviors during a round-trip driving across the first-order point. As result, we find that out-of-equilibrium scaling behaviors are still observed (even after several passages across the FOT), although the associated scaling functions develops a dependence on the details of the driving protocol at the inversion time. \\

The paper is organized as follows. In Sec.~\ref{sec:model}, we introduce the model and briefly recall its phase diagram. Our focus will be on the FOT line and on the finite-size scaling behavior across it. Sec.~\ref{sec:protocol} sets our notation for the driving protocol across the quantum FOT. In particular, we shall consider the case of a linear variation of the driving parameter with time scale $t_s$. In Sec.~\ref{sec:OFSS-theory}, we present the out-of-equilibrium scaling hypothesis arising in the vicinity of the FOT when $L\to \infty$ and $t_s\to \infty$. Numerical results for the many-body system are shown to support the validity of the underlying scaling theory. In Sec.~\ref{sec:effective-description}, we develop an effective description of the quantum spin chain obtained by projecting the many-body Hilbert space onto the subspace spanned by the lowest energy levels competing across the FOT. With this approximation, we determine an analytical expression of the out-of-equilibrium scaling functions. In Sec.~\ref{sec:Floquet}, we extend our analysis to the case of periodic driving across the FOT. Finally, in Sec.~\ref{sec:conclusion} we provide a short summary of our results and draw some conclusions. Appendices \ref{app:eq-FSS-func} and \ref{app:OFSS-func} contain the details on the analytical calculation of the scaling functions at and out of equilibrium, respectively.

\section{The model and the FOT}\label{sec:model}
As a prototypical quantum many-body system displaying a FOT, we shall consider the one-dimensional Ising model in a tilted magnetic field, whose Hamiltonian reads
{\be\label{eq:model}
\hat{H}(h_\perp,h_\parallel)=-J\sum_{j=1}^{L-1} \hat\sigma^{(3)}_j\hat\sigma^{(3)}_{j+1} -\sum_{j=1}^L (h_\perp\hat\sigma^{(1)}_j + h_\parallel \hat\sigma^{(3)}_j).
\ee}
Here, $L$ is the system size, $\hat\sigma _j^{(k=1,2,3)}$ denotes standard Pauli operators acting on site $j$, $h_\perp$ (resp.~$h_\parallel$) is the transverse (resp.~longitudinal) component of the magnetic field and $J=1$ is the overall energy scale which is set to one from thereafter.

 We set open boundary conditions (OBC) for the spin chain~\eqref{eq:model}. Notice that other choices of {\it neutral} boundary conditions (i.e., not favoring any particular phase) --such as periodic boundary conditions-- will not alter our discussion below. The case of non-neutral types of boundary is discussed in Refs.~\cite{fontana2019scaling,pelissetto2020scaling,panagopoulos2018dynamic,campostrini2015quantum,pelissetto2018finite,rossini2018ground}. \\

\begin{figure}[t]
\centering
\includegraphics[width=0.5\textwidth]{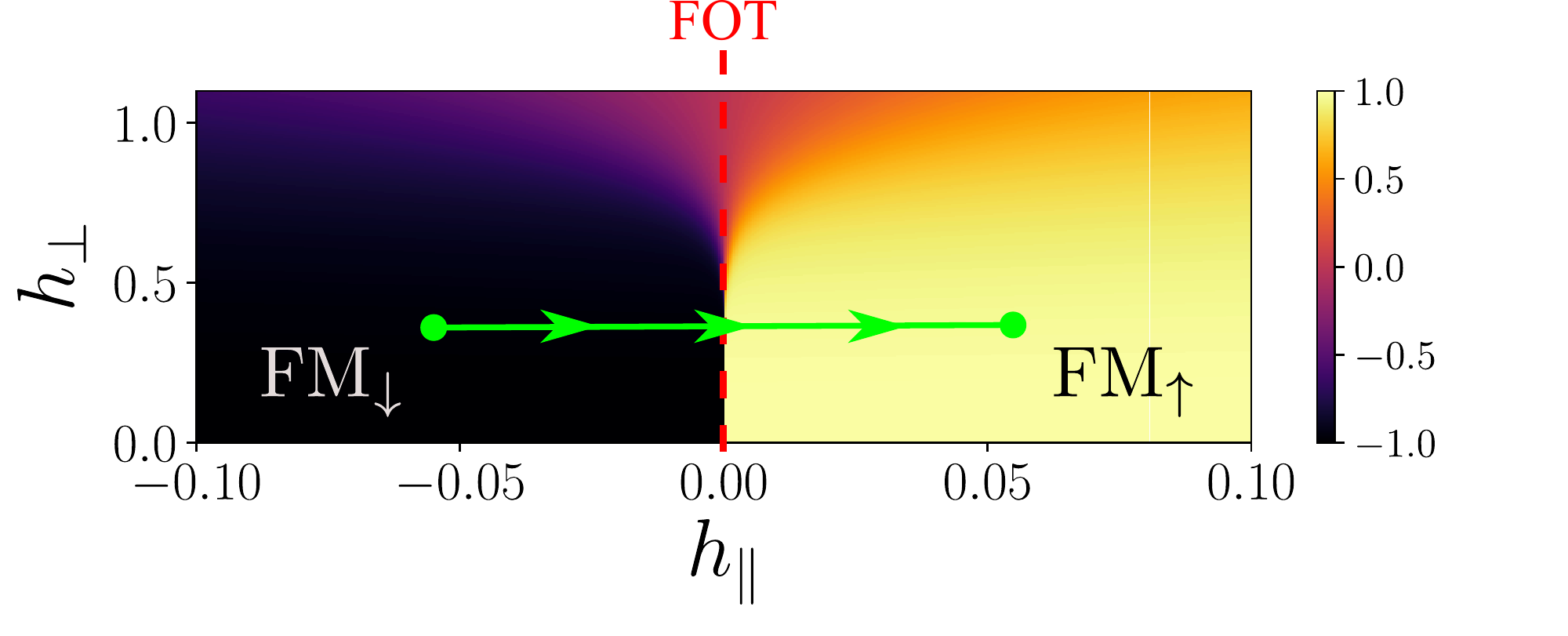}
\caption{Illustration of the phase diagram of the model \eqref{eq:model}~---~For $h_\perp<1$, the ground state is ferromagnetic, i.e., it features a non-vanishing longitudinal magnetization (color map). This ferromagnet is aligned with the direction of $h_\parallel$ (FM$_\downarrow$ and FM$_\uparrow$ phases in the figure). A line of quantum FOTs at $h_\parallel=0$ separate these two ferromagnetic phases (dashed vertical line). The protocol that will be discussed is a passage across the quantum FOT at fixed $h_\perp<1$ (green line).}\label{fig:phase-diag}
\end{figure}

We briefly recall the phase diagram of the model \eqref{eq:model}, depicted in Fig.~\ref{fig:phase-diag}. At $(h_\perp,h_\parallel)=(1,0)$, the Ising model develops a critical behavior belonging to the 2D Ising universality class, see e.g. \cite{S99}. For $h_\parallel\neq 0$ instead, the system is always gapped. Our focus is on the ferromagnetic phase $h_\perp<1$, where the model undergoes a quantum FOT at $h_\parallel=0$. Across this FOT point, the model remains non critical and thus displays exponential decay of correlation functions. Nevertheless, it has been shown that finite-size scaling (FSS) behaviors arise in the limit $L\to\infty$, $h_\parallel\to0^\pm$ \cite{P90}, as argued below.\\

\begin{figure}[t]
\includegraphics[width=7cm]{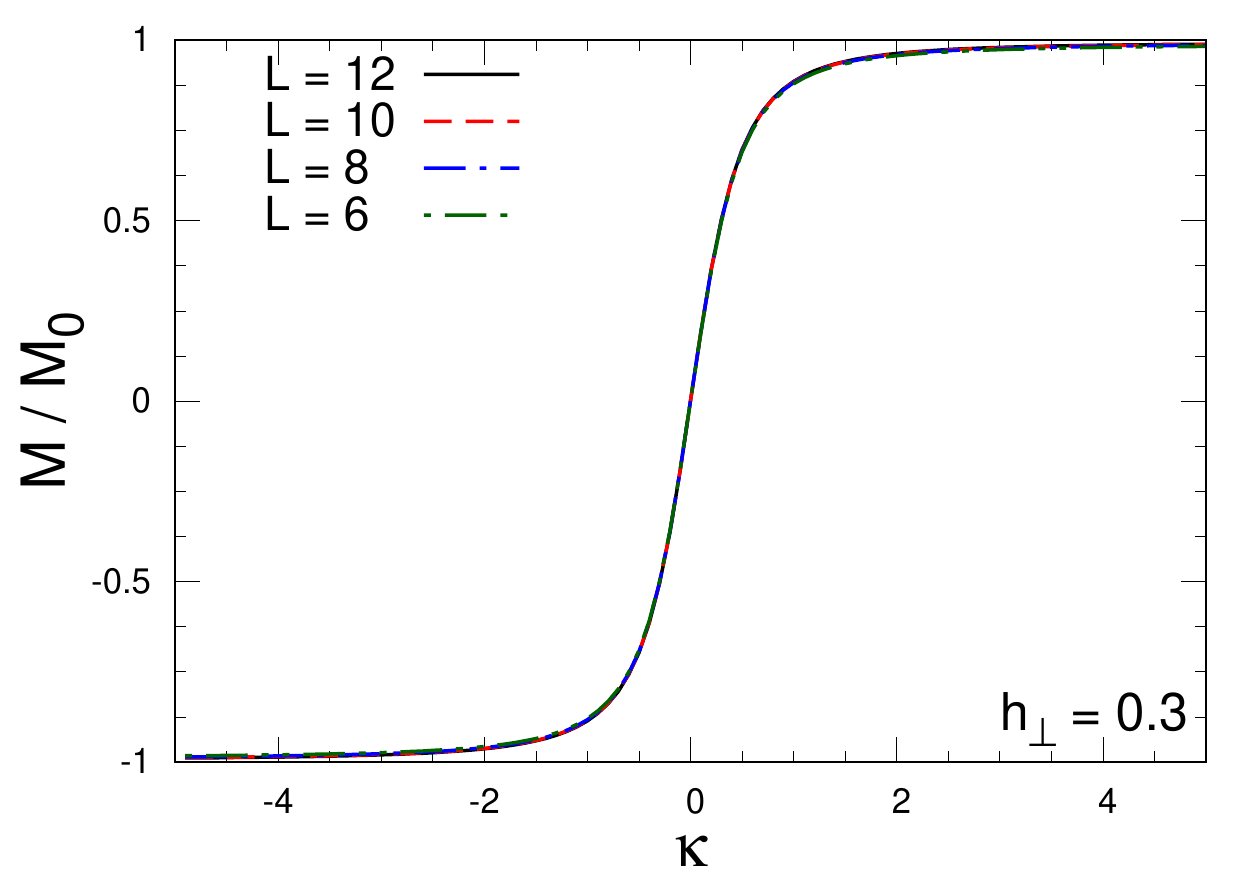}\\
\caption{FSS of the longitudinal magnetization in Eq.~\eqref{eq:eq-scalingM}~--~$M/M_0$ shown as function of the scaling variable $\kappa$ for different system sizes up to $L=12$ and $h_\perp=0.3$. The data collapse to the scaling function $f_M$ in Eq.~\eqref{eq:eq-scaling-func}.}\label{fig:eq-FSS-M}
\includegraphics[width=7cm]{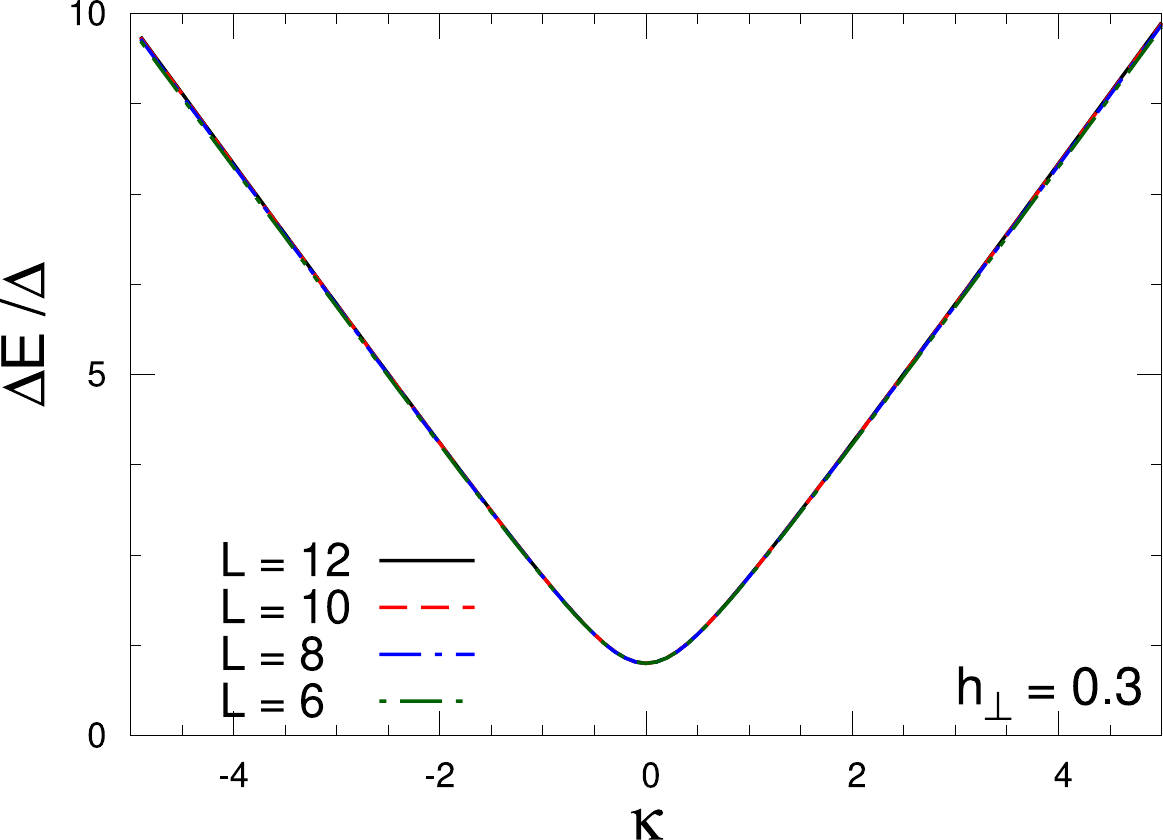}
\caption{FSS of the energy gap in Eq.~\eqref{eq:eq-scalingE}~--~$\Delta E/\Delta$ shown as function of the scaling variable $\kappa$ for different system sizes up to $L=12$ and $h_\perp=0.3$. The data collapse to the scaling function $f_E$ in Eq.~\eqref{eq:eq-scaling-func}.}\label{fig:eq-FSS-E}
\end{figure}

For $h_\parallel=0$, the model \eqref{eq:model} features a level crossing of the two lowest-energy states in the
infinite-volume limit, separated by an exponentially closing energy gap for $L\to\infty$ \cite{cabrera1987role}. For OBC, this is
\be\label{eq:gap}
\Delta(h_\perp,L)= 2h_\perp^L(1-h_\perp^2) \,\Bigr[ 1+{\cal O}(h_\perp^{2L})\Bigr].
\ee 
On the other hand, the presence of a small longitudinal magnetic field $|h_\parallel|\ll 1$ induces a Zeeman-like gap in energy between the two lowest levels, thus introducing another symmetry-breaking mechanism of the aforementioned degeneracy. The latter can be estimated using standard perturbation theory in $h_\parallel$ as \cite{campostrini2014finite}
\be\label{eq:zeeman}
{\cal E}(h_\perp,h_\parallel,L)\overset{h_\parallel\to0}{\simeq} 2h_\parallel \sum_{j=1}^L \left|\langle\hat\sigma^{(3)}_j\rangle\right| \simeq 2h_\parallel L M_0(h_\perp),
\ee
where we approximated the longitudinal magnetization with its value $M_0=(1-h_\perp^2)^{1/8}$ attained when $h_\parallel=0$ and $L=\infty$. \\
These two effects \eqref{eq:gap} and \eqref{eq:zeeman} do compete across a quantum FOT (i.e., when $h_\parallel\to0^\pm$, $L\to\infty$), giving rise to a FSS behavior controlled by the scaling variable $\kappa={\cal E}(h_\perp,h_\parallel,L)/\Delta(h_\perp,L)$ \cite{campostrini2015finite, P90}. For instance, the longitudinal magnetization
\be
M(h_\perp,h_\parallel,L)=L^{-1}\sum_{j=1}^L \langle\hat\sigma^{(3)}_j\rangle
\ee
satisfies the FSS for $h_\parallel \to0^\pm$, $L\to \infty$ \cite{campostrini2015finite, Nienhuis75}
\be\label{eq:eq-scalingM}
M(h_\perp,h_\parallel,L)\sim M_0(h_\perp) \ {f}_M(\kappa).
\ee
Similarly, the energy gap between the two lowest levels $\Delta E(h_\perp,h_\parallel,L)=E_1-E_0$ obeys the FSS
\be\label{eq:eq-scalingE}
\Delta E(h_\perp,h_\parallel,L) \sim \Delta(h_\perp,L) \ {f}_E(\kappa).
\ee
Note that the ground-state proprieties of the model \eqref{eq:model} across the quantum FOT are entirely controlled by the competition of the two quasi-degenerate vacua. This enables the analytical calculation of the scaling functions ${ f}_M,{ f}_E$ in \eqref{eq:eq-scalingM} and \eqref{eq:eq-scalingE} using an effective two-level description (see Appendix~\ref{app:eq-FSS-func} and Refs.\cite{campostrini2014finite, PrivmanFisher85, Fisher82, FisherPrivman85} for details). The result is 
\be\label{eq:eq-scaling-func}
{ f}_M(\kappa)=\frac{\kappa}{\sqrt{1+\kappa^2}}; \qquad { f}_E(\kappa)=\sqrt{1+\kappa^2}.
\ee
The FSS behavior of Eqs.~\eqref{eq:eq-scalingM} and \eqref{eq:eq-scalingE} is shown in Figs.~\ref{fig:eq-FSS-M} and \ref{fig:eq-FSS-E}.\\

In what follows, we investigate the out-of-equilibrium finite-size scaling (OFSS) behaviors arising due to a slowly-varying time-dependent longitudinal magnetic field $h_\parallel$ that drives the system across the quantum FOT at fixed $h_\perp<1$. \\
We discuss the case of single- and round-trip passage through the transition, and we comment on the validity of the two-level effective description of the many-body system during the nonequilibrium dynamics. 

\section{Driving protocol}\label{sec:protocol}
In our setting, the out-of-equilibrium dynamics is generated at fixed $h_\perp<1$ by varying $h_\parallel$ in the Ising Hamiltonian \eqref{eq:model} as a linear ramp in time~\footnote{{The case of a non-linear driving $h_\parallel(t)=(t/t_s)^r$ with odd exponent $r$ can be obtained straightforwardly, see e.g.~Ref.~\cite{dziarmaga2010dynamics}.}}
\be\label{eq:ramp}
h_\parallel(t)=t/t_s,
\ee
where $t_s$ is a characteristic time scale. With this convention, the model is prepared at $t_i<0$ in the many-body ground state $\ket{\Psi(t_i)}=\ket{\Psi_0[h_\parallel(t_i)]}$ corresponding to the initial value of longitudinal field $h_\parallel=t_i/t_s$. At times $t>t_i$, the system evolves unitarily with time-dependent Ising Hamiltonian 
\be\label{eq:schrodinger}
i\de_t \ket{\Psi(t)}= \hat{H}(h_\perp<1, t/t_s)\ket{\Psi(t)},
\ee
crossing the quantum FOT when $t=0$. A single-passage protocol stops at a final time $t_f>0$ while a round-trip protocol is implemented by inverting the ramp \eqref{eq:ramp} at $t=t_f$ so that we can drive the system back to the value $h_\parallel(t_i)$ in a time window $2(t_f-t_i)$;  see Fig.~\ref{fig:protocol} for an illustration.\\
\section{Out-of-equilibrium FSS at FOT}\label{sec:OFSS-theory}
Below, we specify to the case of slow drivings, $t_s\to\infty$. For a single passage and at the critical point ($h_\perp=1$), this class of protocols would correspond to the standard Kibble-Zurek setup, whose OFSS has been extensively discussed in literature, see e.g. Refs.~\cite{dziarmaga2005dynamics,polkovnikov2005universal,zurek2005dynamics,collura2010critical,liu2020kibble,cui2020experimentally,mukherjee2020universal,zeng2023universal,dutta2016anti,sadhukhan2020sonic,ulm2013observation,weiler2008spontaneous,pyka2013symmetry,navon2016emergence} and \cite{dziarmaga2010dynamics,rossini2021coherent} for reviews. {We also mention recent works, e.g.~\cite{delcampo2020full,bando2020probing,gomez2022role}, on the complete probability distribution of topological defects}. 
\\

Analogously, across the quantum FOT $(h_\perp<1,h_\parallel=0)$, one can formulate an OFSS ansatz as the limit  $L\to\infty$, $u\equiv  {t_s L^{-1} M_0^{-1}}\to \infty$ with fixed scaling variables:
\begin{align}\label{eq:OFSS-scaling-var}
&\tau=t/\sqrt{u};\\
\label{eq:OFSS-scaling-var2}
& \upsilon=u \ \Delta(h_\perp, L)^2 ,
\end{align}
where the longitudinal magnetization is expected to scale as
\begin{equation}\label{eq:OFSS-M}
M(h_\perp,t,t_s,L)\sim M_0(h_\perp) \  {\cal F}_M(\tau,\upsilon)
\end{equation}
with OFSS function ${\cal F}_M$. \\
\begin{figure}[t]
\centering
\includegraphics[width=0.45\textwidth]{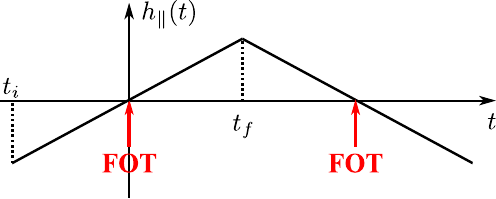}
\caption{Illustration of the driving protocol: the longitudinal magnetic field $h_\parallel(t)$ is varied as a linear ramp \eqref{eq:ramp} from a time $t_i<0$ to a time $t_f>0$ with slope $t_s^{-1}$. With this convention, the system is driven across the quantum FOT at $t=0$. For a round-trip passage, the ramp is inverted at $t=t_f$ and the system crosses the quantum FOT also at $t=2t_f$. }\label{fig:protocol}
\end{figure}
In Eq.~\eqref{eq:OFSS-scaling-var}, $\tau$ is a rescaled time and $t_{\rm KZ}=\sqrt{u}$ plays the role of a Kibble-Zurek time, as commented in the following section. Similar relations arise at classical FOT, see e.g. Refs.~\cite{pelissetto2016off,scopa2018dynamical}. In this sense, we expect to assist to a breakdown of adiabaticity when $|\tau|\lesssim 1$. To quantitatively probe this effect during the driving, it is useful to introduce the {\it adiabaticity function}
\begin{equation}
	\label{eq:adiab}
	A(h_\perp,t,t_s,L) = \Big| \braket{\Psi_0\bigr[h_\parallel(t)\bigr]}{\Psi(t)} \Big| \,\,,
\end{equation}
defined as the modulus of the overlap coefficient between the time-evolved wavefunction $\ket{\Psi(t)}$ and the instantaneous ground state $\ket{\Psi_0[h_\parallel(t)]}$ of the time-dependent Hamiltonian $\hat{H}(h_\perp,t/t_s)$. Initially, $A(t_i)=1$ by construction, and it deviates from one when the adiabatic approximation of the state $\ket{\Psi(t)}$ breaks down near the quantum FOT point. In the OFSS limit,
\be\label{eq:OFSS-A}
A(h_\perp,t,t_s,L)\sim {\cal F}_A(\tau,\upsilon).
\ee
Finally, notice that the scaling variable $\kappa$ (characterizing the equilibrium FSS) is obtained from \eqref{eq:OFSS-scaling-var} and \eqref{eq:OFSS-scaling-var2} as
\be\label{eq:kappa-ofss}
\kappa = {\frac{2\tau}{\sqrt{\upsilon}}}.
\ee
It is then easy to see that the energy gap 
\be
\Delta E(h_\perp,t,t_s,L)\sim \Delta(h_\perp,L) \ f_E\left( {\frac{2\tau}{\sqrt{\upsilon}}}\right)
\ee
 in the OFSS limit.\\

\begin{figure}[t]
\includegraphics[width=7cm]{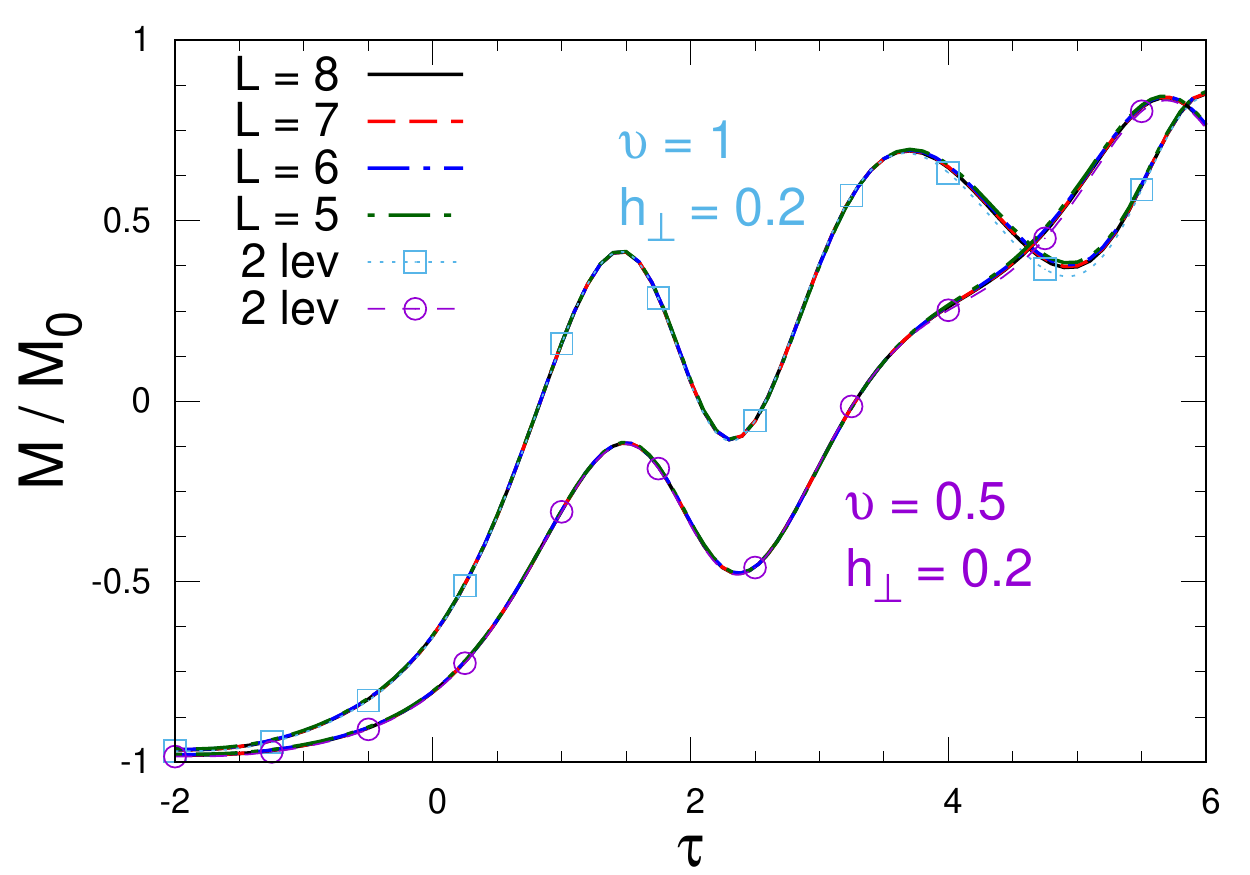}
\caption{OFSS of the longitudinal magnetization in Eq.~\eqref{eq:OFSS-M}~---~$M(t)/M_0$ shown as a function of the rescaled time $\tau$ during a round-trip protocol with $|\tau _i|=\tau_f= 2$ (FOTs at $\tau=0,4$). We show different values of $\upsilon$ and $h_\perp$ (different curves) and we vary the system sizes up to $L = 8$. {In the plot legend, `2 lev' refers to the scaling functions ${\cal F}_M(\tau,\upsilon)$ obtained using the effective two-level description discussed in Sec.~\ref{sec:2-lev}.}.}
\label{LongMagn}
\end{figure}

\begin{figure}[t]
		\includegraphics[width=7cm]{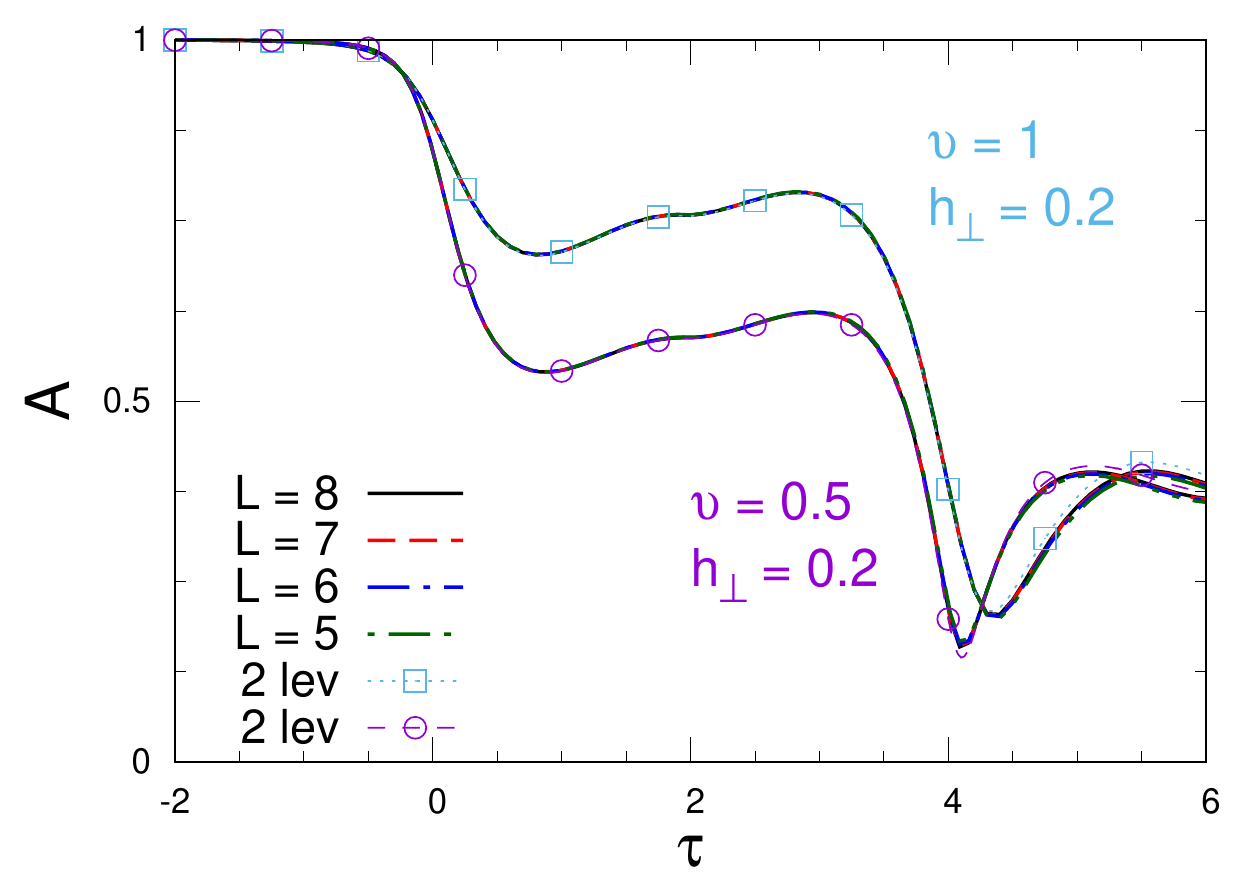}
		\caption{OFSS  of the adiabaticity function in Eq.~\eqref{eq:OFSS-A}~---~$A(t)$ shown as a function of the rescaled time $\tau$ during a round-trip protocol with $|\tau_i|=\tau_f =2$ (FOTs at $\tau=0,4$). We show different values of $\upsilon $ and $h_\perp$ (different curves) and we vary the system size up to $L = 8$.  {In the plot legend, `2 lev' refers to the scaling functions ${\cal F}_A(\tau,\upsilon)$ obtained using the effective two-level description discussed in Sec.~\ref{sec:2-lev}.}}
		\label{AtripAt2u05g02}
	\end{figure}

\begin{figure}[t]
\includegraphics[width=7cm]{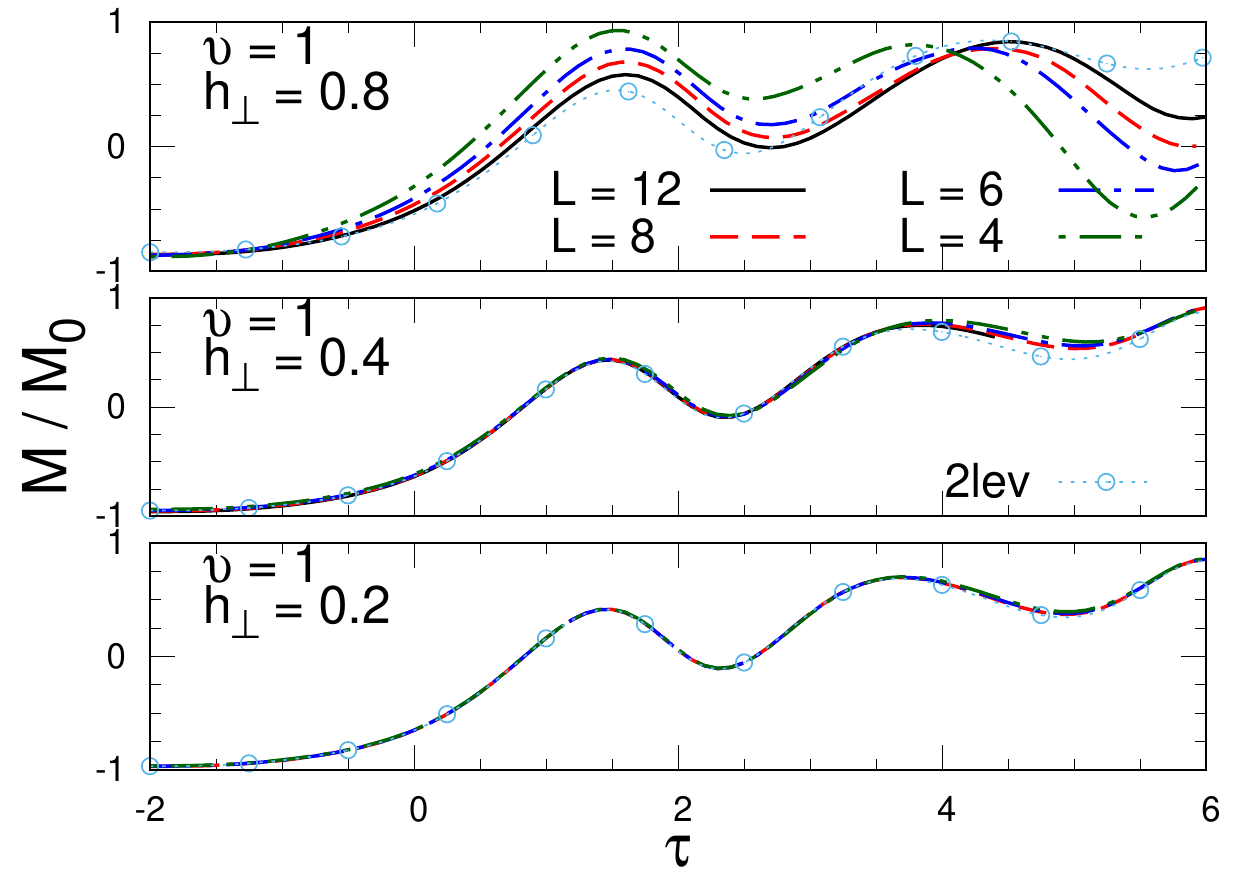}
\includegraphics[width=7cm]{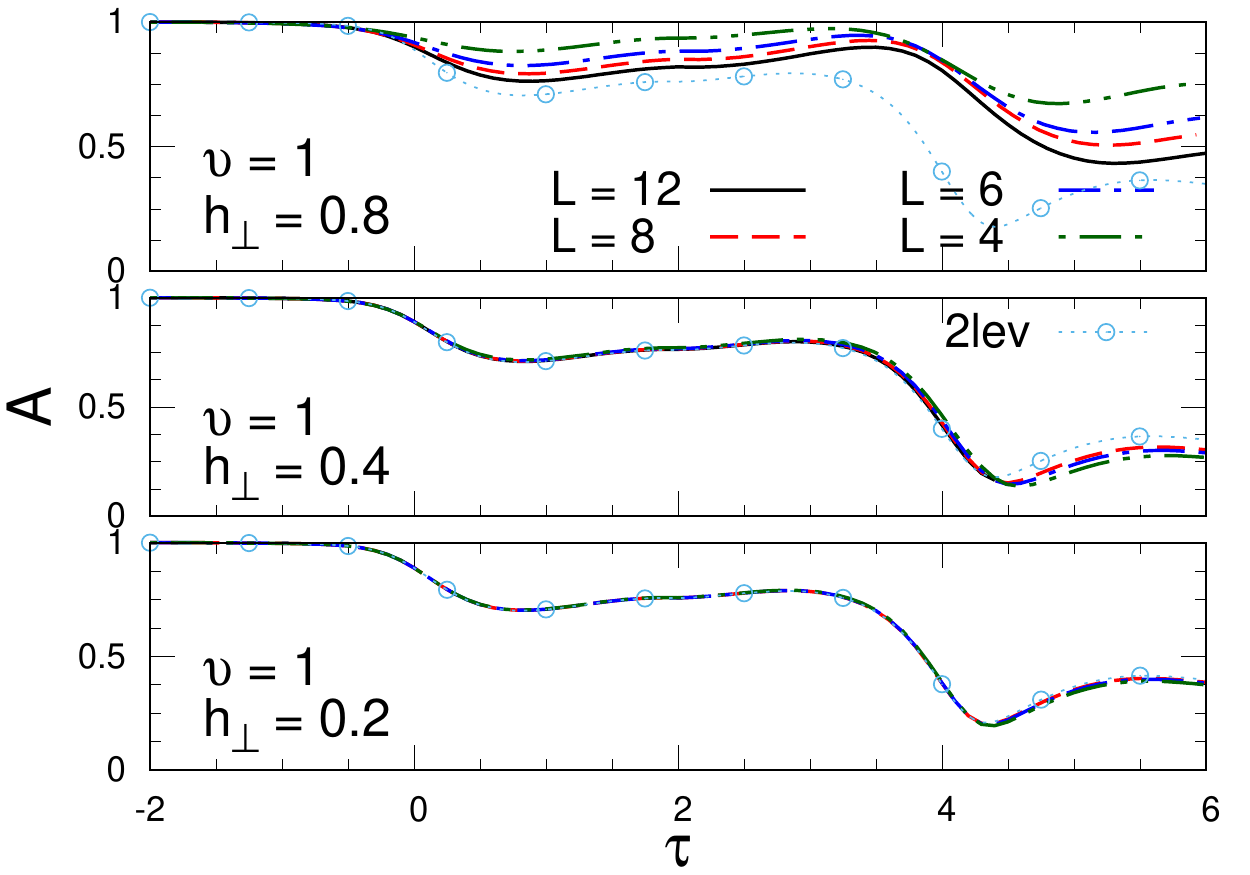}
\caption{{Convergence to the OFSS regime~--~Top: Longitudinal magnetization $M(t)/M_0$; Bottom: adiabaticity function $A(t)$, both shown for different $L$ and $h_\perp$ as function of the rescaled time $\tau$ during a round-trip protocol with $|\tau_i|=\tau_f=2$ and $\upsilon=1$. The symbols show the scaling functions ${\cal F}_M$, ${\cal F}_A$ obtained using the effective two-level description discussed in Sec.~\ref{sec:2-lev}.} }
\label{fig:convergence}
\end{figure}

In Figs.~\ref{LongMagn} and \ref{AtripAt2u05g02}, we show the results for the OFSS of the longitudinal magnetization (Eq.~\eqref{eq:OFSS-M}) and of the adiabaticity function (Eq.~\eqref{eq:OFSS-A}). The numerical data is obtained by performing exact diagonalization of the spin chain \eqref{eq:model} and Runge-Kutta methods for time evolution. Notice that despite our data are obtained for spin chains of modest system sizes, the convergence to the thermodynamic limit is controlled by the interplay of $h_\perp$ and $L$, and thus it can be reached already for modest system sizes when $h_\perp\ll 1$. For $h_\perp$ closer to one, larger values of $L$ are required to observe OFSS, see Refs.~\cite{campostrini2014finite,campostrini2015finite}.{The convergence to the OFSS regime is shown in Fig.~\ref{fig:convergence} for different values of $L$ and $h_\perp$.}\\

The OFSS for a single passage across the quantum FOT has been discussed e.g. in Refs.~\cite{pelissetto2018out,sinha2021nonadiabatic,coulamy2017dynamics,shimizu2018dynamics,qiu2020observation,rossini2020dynamic}. Here, we show that such scaling behavior remains valid for a round-trip passage. It is important to remark that the OFSS during a round-trip protocol cannot be inferred from the observed OFSS for a single passage. Indeed,  while the first passage is responsible for the formation of excitations from the adiabatic ground state in the Kibble-Zurek sense, for $t>t_f$ the system is found in a nonequilibrium state before approaching the transition for the second time. Therefore, standard Kibble-Zurek arguments do not apply.\\

\begin{figure}[t]
\includegraphics[width=7cm]{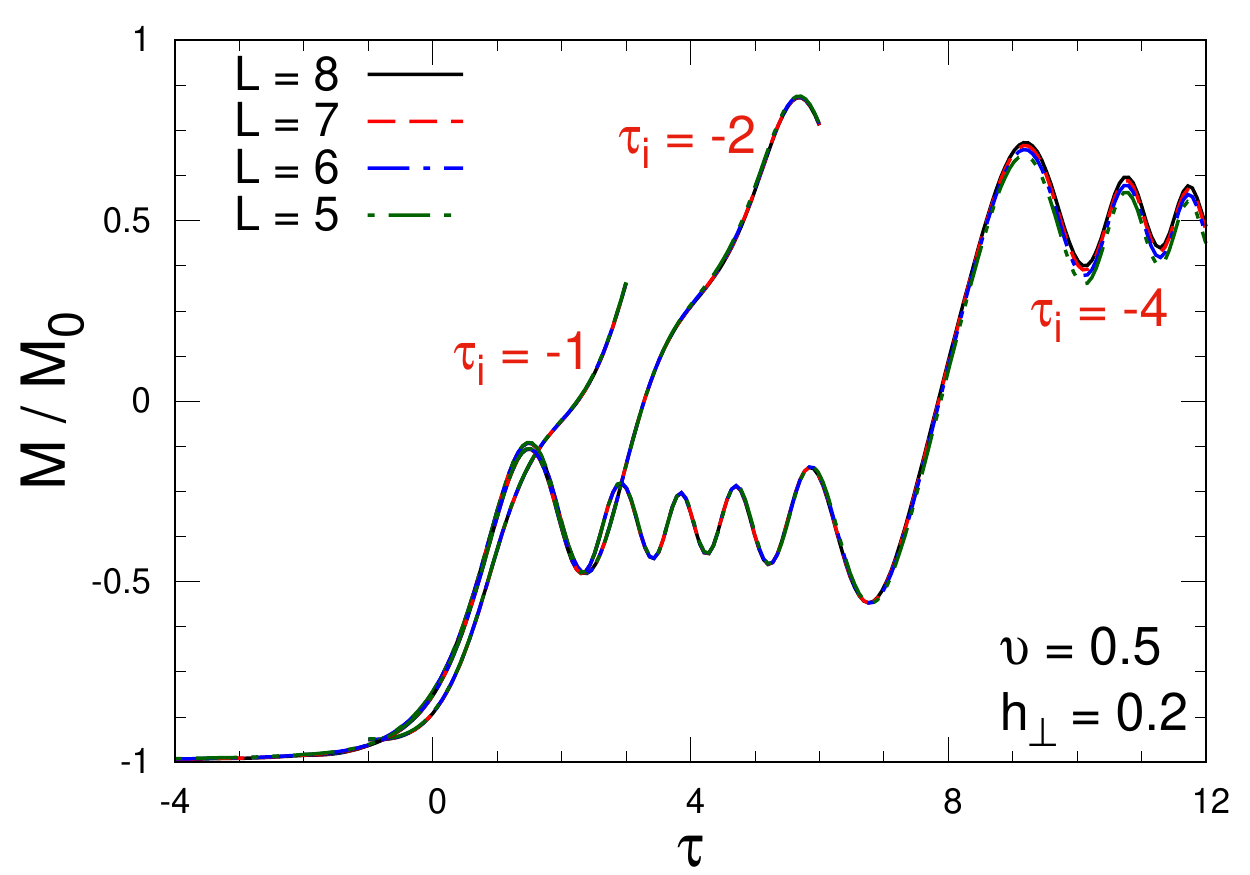}
\caption{Quasi-universality of the OFSS behavior of $M/M_0$ during a round trip protocol. We show the data collapse as function of $\tau$ at fixed $\upsilon=0.5$, $h_\perp=0.2$, and varying the system size up to $L=8$. Different curves refer to different values of $|\tau_i|=\tau_f$ (see legend).}\label{fig:dependence-ti-M}
\end{figure}

\begin{figure}[t]
\includegraphics[width=7cm]{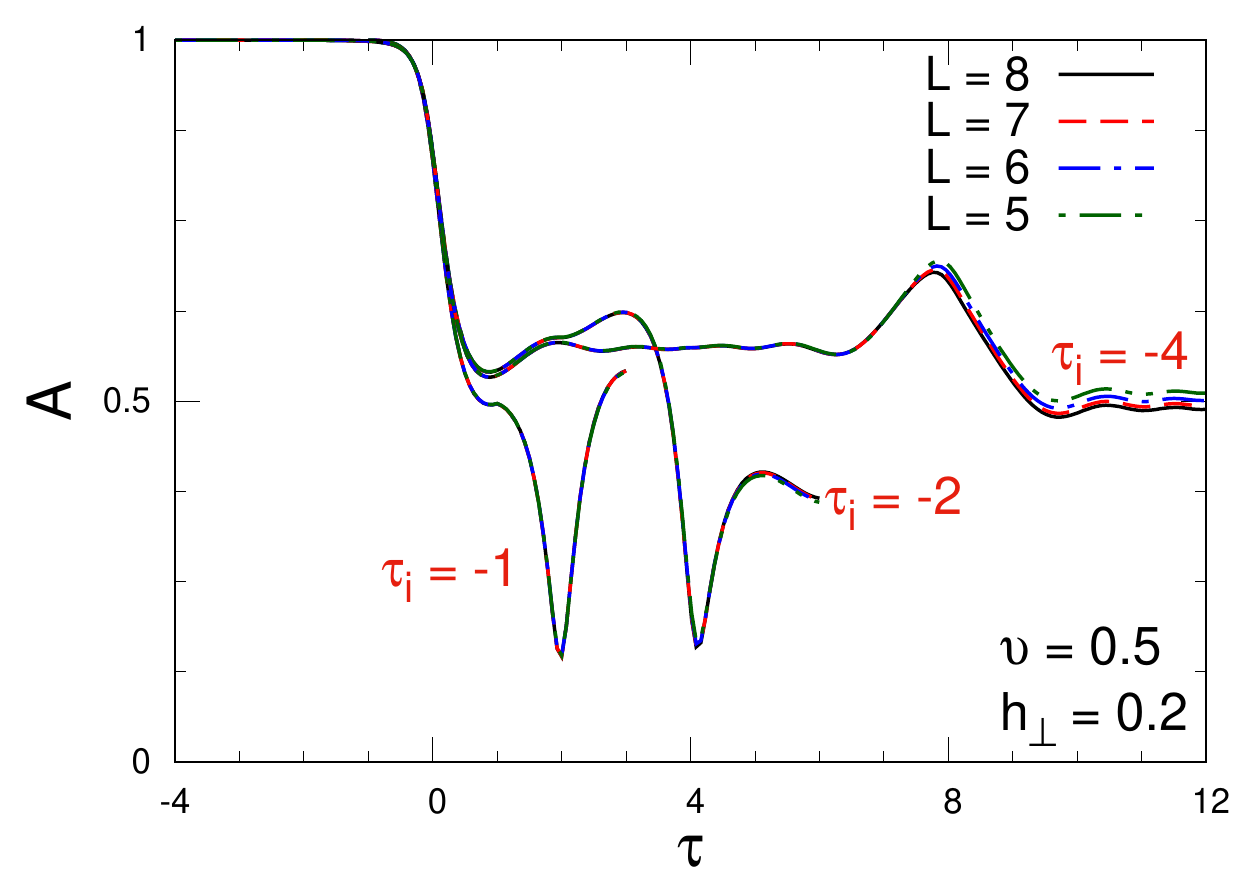}
\caption{Quasi-universality of the OFSS behavior of $A$ during a round trip protocol. We show the data collapse as function of $\tau$ at fixed $\upsilon=0.5$, $h_\perp=0.2$, and varying the system size up to $L=8$. Different curves refer to different values of $|\tau_i|=\tau_f$ (see legend).}\label{fig:dependence-ti-A}
\end{figure}

Interestingly, the OFSS during the round-trip protocol shows a dependence on the initial condition $\tau_i$, see Figs.~\ref{fig:dependence-ti-M} and \ref{fig:dependence-ti-A}. Such feature arises due to the magnetic field inversion occurring at $t=t_f$, i.e., when the system is already far from equilibrium and thus unable to wash out such information during the unitary evolution for $t>t_f$. This is a qualitative difference with respect to the single passage protocol. Given the dependence on $\tau_i$, we refer to the OFSS functions ${\cal F}_M$ and ${\cal F}_A$  for the round trip as {\it quasi-universal}.\\

In the following sections, we derive an effective description of the many-body system during the driving protocol \eqref{eq:ramp}. In this way, we determine analytical expressions for the quasi-universal OFSS functions characterizing the round-trip protocol.
 
\section{Effective description in the OFSS regime}\label{sec:effective-description}
Let us consider the following formal expansion of the many-body wavefunction
\be\label{eq:expansion}
\ket{\Psi(t)}=\sum_{n=0}^{2^L-1} C_n(t) e^{-i\vartheta_n(t)} \ket{\psi_n(t)}
\ee
where $\ket{\psi_n(t)}$ is the instantaneous eigenbasis of the Hamiltonian at fixed time $t$ satisfying
\be\label{eq:eigenproblem}
\hat{H}(t)\ket{\psi_n(t)} = E_n(t) \ket{\psi_n(t)},
\ee
and
\be
\vartheta_n(t)=\int_{t_0}^t ds \ E_n(s)
\ee
is the associated dynamical phase. Using \eqref{eq:expansion}, the time evolution generated by the Schr\"odinger equation \eqref{eq:schrodinger} can be reduced to the set of equations for the overlap coefficients
\be\label{eq:MB-coef-evo}
\frac{d C_k}{dt}= -\sum_n C_n(t)
\bra{\psi_k(t)}\ket{\frac{\partial\psi_n(t)}{\partial t}}
e^{i(\vartheta_n(t)-\vartheta_k(t))},
\ee
which is solved imposing the initial condition $C_k(t_i)=\delta_{0,k}$. Differentiating Eq.~\eqref{eq:eigenproblem}, one obtains
\be\begin{split}
&\bra{\psi_m(t)}\de_t \hat{H} \ket{\psi_n(t)}\\
&+(E_m(t)-E_n(t))\bra{\psi_m(t)}\ket{\frac{\de \psi_n(t)}{\de t}} =
\frac{dE_m}{dt} \delta_{m,n}
\end{split}\ee
from which one can write Eq.~\eqref{eq:MB-coef-evo} as
\be\begin{split}\label{eq:coeff-2}
\frac{d C_k(t)}{dt}&= -\bra{\psi_k(t)}\ket{\frac{\de \psi_k(t)}{\de t}}
C_k(t)\\[2pt]
&+ \sum_{n\neq k} C_n(t) \frac{\bra{\psi_k(t)}\de_t\hat{H}
\ket{\psi_n(t)}}{(E_n(t)-E_k(t))} e^{i(\vartheta_n(t)-\vartheta_k(t))}.
\end{split}\ee
This formal expression for the coefficients is exact. Specifying it to the Ising Hamiltonian \eqref{eq:model} with ramp \eqref{eq:ramp}, we have
\be
\frac{\bra{\psi_k(t)}\de_t\hat{H}
\ket{\psi_n(t)}}{(E_n(t)-E_k(t))}=\frac{M_{k,n}(t)}{t_s L^{-1} (E_n(t)-E_k(t))},
\ee
where $M_{k,n}(t)=L^{-1}\bra{\psi_k(t)}\sum_{j=1}^L
{\hat\sigma^{(3)}_j}\ket{\psi_n(t)}$.\\

 The adiabatic limit corresponds to the
limit $t_s\to \infty$ (regardless the value of $L$). In such limit the
off-diagonal terms in Eq.~\eqref{eq:coeff-2} vanish and the system remains in
the adiabatic ground state with $h_\parallel(t)$ at any time $t$. Conversely, in the OFSS limit considered, a breakdown of the adiabatic approximation is observed (cf~Fig.~\ref{AtripAt2u05g02}). In
particular, it is easy to show that the contribution coming from the first
excited level is non-negligible since
\be\label{eq:energy-gap-01}
t_s  L^{-1}(E_1(t)-E_0(t))\sim \sqrt{u \ \upsilon} \ {M_0} \ { f}_E(\kappa)
\ee
and $\upsilon$ is fixed in the OFSS regime. This means that the time scale of
the transition to the first excited level is of the order of $t_{\rm KZ}=\sqrt{u}$, and it starts to be populated at rescaled times $\tau\sim{\cal O}(1)$, in agreement with
 the OFSS arguments above and with the numerical results of Figs.~\ref{LongMagn} and \ref{AtripAt2u05g02}.\\

Higher energy levels are not expected to contribute to the early stages of the nonequilibrium dynamics as we now argue. In the OFSS regime, we can approximate the energy gaps for $n\geq 2$ as
\be
E_n(t)-E_0(t)\approx (E_n-E_0)_{\vert_{h_\parallel=0}}
\ee
since the presence of a weak longitudinal field do not significantly alter higher excited levels. Considering that,
\be
(E_n-E_0)_{\vert_{h_\parallel=0}}\geq (E_2-E_0)\vert_{h_\parallel=0}=2(1-h_\perp)+{\cal O}(L^{-2})
\ee
for OBC \cite{cabrera1987role}, and that $E_n-E_1\simeq E_n-E_0$ up to exponentially small corrections in $L$, we conclude that OFSS regime is restricted within the Hilbert space spanned by the two lowest levels of the Hamiltonian, up to large-time corrections that occur at rescaled times of order $\tau\sim
{\cal}O(t_{\rm KZ})$.\\

In order to further check the validity of this argument, we introduce the quantity
\be
B(h_\perp,t,t_s,L)= \Big| \braket{\Psi_1\bigr[h_\parallel(t)\bigr]}{\Psi(t)} \Big|,
\end{equation}
where $\ket{\Psi_1\bigr[h_\parallel(t)\bigr]}$ is the instantaneous first excited state associated with 
the Hamiltonian $\hat{H}(h_\perp,t/t_s)$ in \eqref{eq:model}. Alongside with the adiabaticity function \eqref{eq:adiab}, the time evolution of $B(t)$ is able to probe the validity of the two-level approximation during the driving. In this sense, introducing {\it totality function}
\begin{equation}
	\label{totalAB}
	T(t) = \sqrt{A^2(t) + B^2(t)},
\end{equation}
the distance variable
\begin{align}
	\label{distance}
	D(t) = 1 - T(t)
\end{align}
gives an estimate of the error arising from the projection of the many-body state $\ket{\Psi(t)}$ onto a two-dimensional Hilbert space spanned by $\bigl\{ \ket{\Psi_0\bigr[h_\parallel(t)\bigr]};\ket{\Psi _1\bigr[h_\parallel(t)\bigr]} \bigl\}\,$. In particular, the closer $D(t)$ is to zero, the more accurate the two-level approximation is.\\
	\begin{figure}[t]
		\includegraphics[width=7cm]{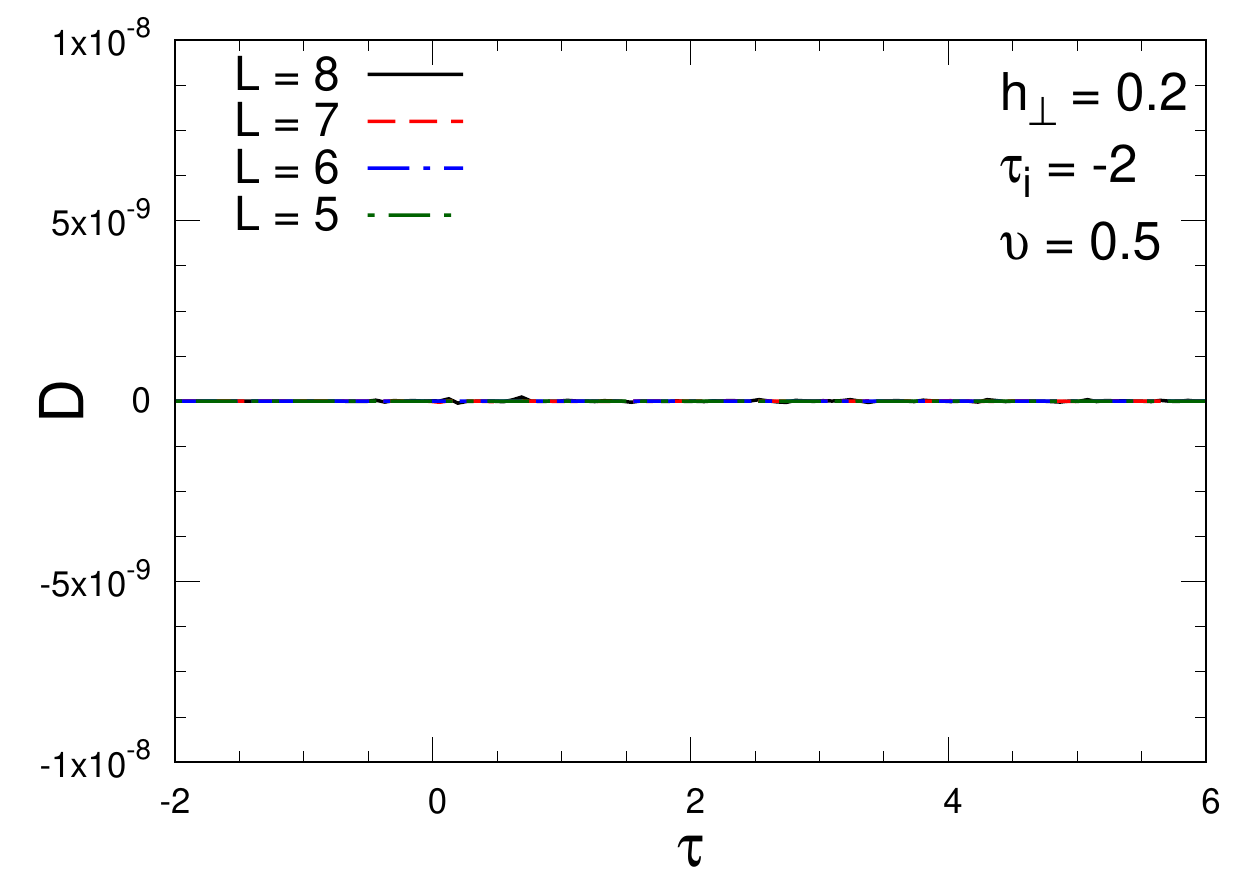}
		\caption{OFSS of the function $D(t)$ as a function of the rescaled time $\tau$ during a round-trip protocol with $|\tau_i|=\tau_f=2$ (FOTs at $\tau=0,4$). We set $\upsilon= 0.5$, $h_\perp = 0.2$ and different system sizes up to $L = 8$.}
		\label{tripDt2u05g02}
		\includegraphics[width=7cm]{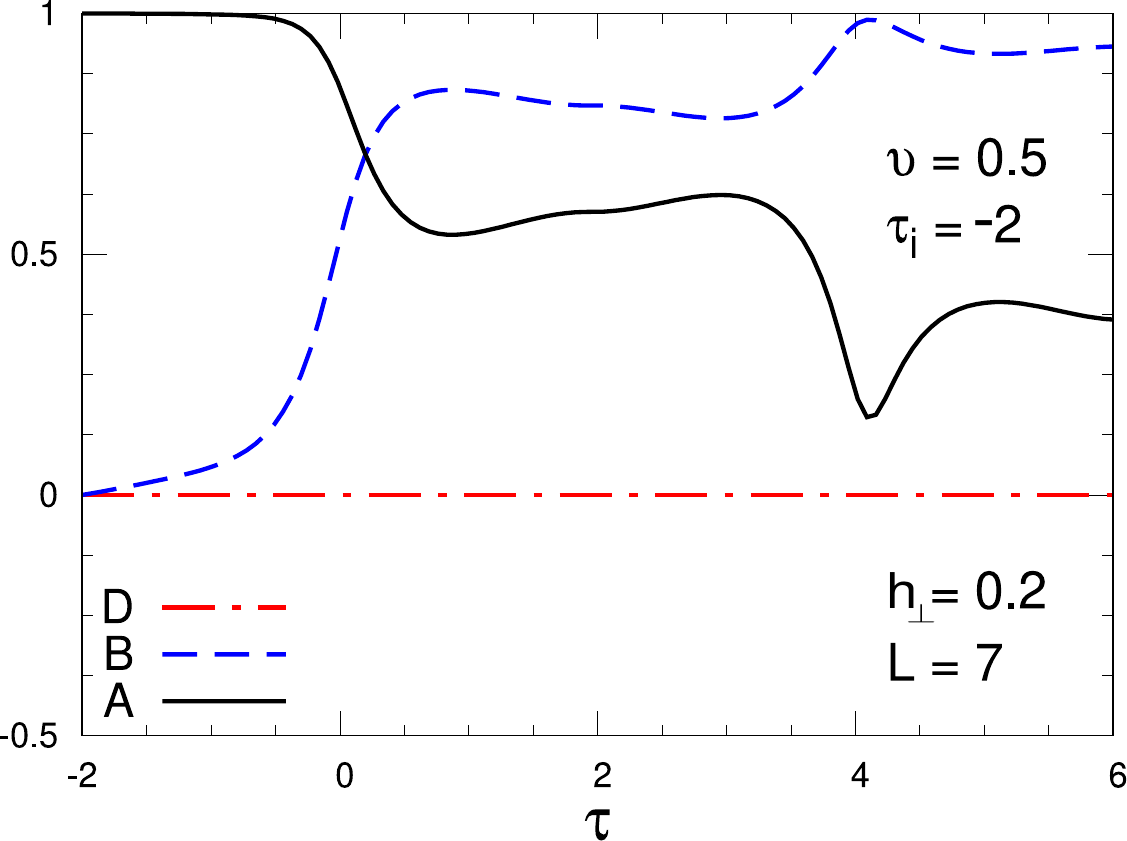}
		\caption{Behavior of $A(t)$, 
		$B(t)$ and $D(t)$ as function of the rescaled time $\tau$ 
		 for a fixed system size $L = 7$ during a round-trip protocol with $|\tau_i|=\tau_f=2$ (FOTs at $\tau=0,4$). We set $\upsilon = 0.5$ and $h_\perp = 0.2$.}
		\label{tripABDt2u05g02L7}
	\end{figure}
The time evolution of $D(t)$ for the driving of Figs.~\ref{LongMagn} and \ref{AtripAt2u05g02} is shown in Fig.~\ref{tripDt2u05g02}. As one can see, the distance $D(t)$ remains extremely close to zero during the whole driving. In Fig.~\ref{tripABDt2u05g02L7}, we show the time evolution of the functions $A(t)$ and $B(t)$, which undergo (as expected) a non-trivial dynamics characterized by a breakdown of adiabaticity close to the quantum FOT.\\

 Motivated by this evidence, in the next section we develop a two-level effective description of the dynamics to determine the OFSS functions ${\cal F}_M$, ${\cal F}_A$ appearing in Eqs.~\eqref{eq:OFSS-M} and \eqref{eq:OFSS-A}.
\subsection{Two-level model}\label{sec:2-lev}
By projecting the Ising Hamiltonian \eqref{eq:model} onto the two-dimensional Hilbert space spanned by  $\bigl\{ \ket{\Psi_0[h_\parallel(t)]};\ket{\Psi _1[h_\parallel(t)]} \bigl\}$, we obtain (up to an unimportant energy shift) the following effective two-level model 
\be\label{eq:eff-Ham}
\hat{H}_\text{eff}(t)=\frac{1}{2}\left(\Delta(h_\perp,L) \ \hat\sigma^{(1)}-{\cal E}(h_\perp,t,t_s,L)\ \hat\sigma^{(3)}\right),
\ee
with ${\cal E}$ given by Eq.~\eqref{eq:zeeman} with longitudinal field \eqref{eq:ramp}. {Here, we choose the basis $\{\ket{\psi_0}, \ket{\psi_1}\}$ such that ${\cal E}\hat\sigma^{(3)}$ is the perturbation induced by $h_\parallel$, and the off-diagonal terms encode the gap \eqref{eq:gap} between the two quasi-degenerate levels at finite sizes \cite{campostrini2014finite}.} {Higher-order perturbative effects in $h_\perp$ generating large bubbles of true vacuum onto the metastable state can be studied using a multilevel effective model, see Ref.~\cite{sinha2021nonadiabatic} for details.  Notice that Eq.~\eqref{eq:eff-Ham} is equivalent to truncate the expansion of the many-body wavefunction in the instantaneous eigenbasis \eqref{eq:expansion} with the two lowest overlap coefficients
\begin{align}
\ket{\Psi(t)}&=C_0(t) \ket{\Psi_0[h_\parallel(t)]}+ C_1(t) \ket{\Psi _1[h_\parallel(t)]} +{\cal O}(t_{\rm KZ}^{-2})
\nonumber\\[4pt]
&=c_0(t) \ket{\psi_0}+ c_1(t) \ket{\psi _1} +{\cal O}(t_{\rm KZ}^{-2}),
\end{align}
since $\ket{\psi_{0,1}}\simeq \ket{\Psi_{0,1}[h_\parallel(t_0)]}$ for $|t_0|\gg 1$. The two sets of coefficients are then related via the rotation
\be
\begin{pmatrix} C_0(t) \\[4pt] C_1(t) \end{pmatrix}= \begin{pmatrix} \cos(\alpha/2) & -\sin(\alpha/2) \\[4pt] \sin(\alpha/2) & \cos(\alpha/2) \end{pmatrix}\begin{pmatrix} c_0(t) \\[4pt] c_1(t) \end{pmatrix}
\ee
with angle $\alpha=\arctan(\frac{\sqrt{\upsilon}}{2\tau})$ (cf Eq.~\eqref{eq:kappa-ofss} and Appendix~\ref{app:eq-FSS-func}).}\\

Under this approximation, Eq.~\eqref{eq:MB-coef-evo} reduces to a finite-time Landau-Zener-St\"uckelberg (LZS) problem in the interval $t\in[t_i,t_f]$ \cite{landau1932theorie, zener1932non}
\be
i\frac{d}{dt} \begin{pmatrix} c_0(t)\\[4pt] c_1(t) \end{pmatrix}=\frac{1}{2}\begin{pmatrix} -{\cal E}(t) & \Delta \\[4pt] \Delta & {\cal E}(t)\end{pmatrix}  \begin{pmatrix} c_0(t)\\[4pt] c_1(t) \end{pmatrix}.
\ee
In terms of the OFSS variables \eqref{eq:OFSS-scaling-var} and \eqref{eq:OFSS-scaling-var2},
\be
i\frac{d}{d\tau} \begin{pmatrix} c_0(\tau,\upsilon)\\[4pt] c_1(\tau,\upsilon) \end{pmatrix}=\begin{pmatrix} -\tau & \frac{\sqrt\upsilon}{2} \\[4pt] \frac{\sqrt\upsilon}{2} & \tau\end{pmatrix}  \begin{pmatrix} c_0(\tau,\upsilon)\\[4pt] c_1(\tau,\upsilon) \end{pmatrix},
\ee
which can be solved imposing that $(c_0(\tau_i,\upsilon)\ ;\ c_1(\tau_i,\upsilon))=(1;0)$, see Appendix~\ref{app:OFSS-func} for details on the calculation. We write the result in terms of the $2\times 2$ Hermitian matrix $U(\tau,\tau_i)$,
\be\label{eq:LZS-single}
\begin{pmatrix} c_0(\tau)\\[4pt] c_1(\tau)\end{pmatrix}= U(\tau,\tau_i) \begin{pmatrix} 1 \\[4pt] 0\end{pmatrix}
\ee
from which the OFSS functions in Eqs.~\eqref{eq:OFSS-M} and \eqref{eq:OFSS-A} during a single-passage protocol are obtained as \cite{pelissetto2018out} 
\begin{align}\label{eq:OFSS-func-M-single}
{\cal F}_M(\tau,\upsilon)&=2|c_1(\tau,\upsilon)|^2-1
\nonumber\\
&=\frac{\upsilon}{4}e^{-\frac{\pi\upsilon}{16}}|\mathscr{D}_{-1+\frac{i\upsilon}{8}}(\sqrt{2}e^{i\frac{3\pi}{4}}\tau)|^2-1
\end{align}
{\begin{align}
\label{eq:OFSS-func-A-single}
&{\cal F}_A(\tau,\upsilon)=|C_0(\tau,\upsilon)|
\nonumber\\
&=e^{-\frac{\pi\upsilon}{32}}\Bigg|\sqrt{\frac{1}{2}+ \frac{|\tau|}{\sqrt{4\tau^2+\upsilon}}} \mathscr{D}_{\frac{i u}{8}}(\sqrt{2}e^{i\frac{3\pi}{4}}\tau) 
\nonumber\\
&\quad -\frac{\sqrt{\upsilon} e^{-\frac{i\pi}{4}}}{2\sqrt{2}} \sqrt{\frac{1}{2}- \frac{|\tau|}{\sqrt{4\tau^2+\upsilon}}} \mathscr{D}_{-1+\frac{i u}{8}}(\sqrt{2}e^{i\frac{3\pi}{4}}\tau)\Bigg|
\end{align}}
and similarly for other quantities. Here, $\mathscr{D}_\nu(z)$ denotes the Parabolic Cylinder function. \\

For the round-trip protocol, we can solve the associated LZS problem in the time window $t\in[t_f,2t_f+|t_i|]$ obtaining:
\be\label{eq:LZS-round-trip}
\begin{pmatrix} c_0(\tau)\\[4pt] c_1(\tau)\end{pmatrix}= \tilde{U}(\tau,\tau_f) U(\tau_f,\tau_i) \begin{pmatrix} 1 \\[4pt] 0\end{pmatrix},
\ee
where $\tilde{U}(\tau,\tau_f)$ is the evolution matrix with inverted time ramp \eqref{eq:ramp}, see Appendix~\ref{app:OFSS-func} for the analytical expression of its elements. From \eqref{eq:LZS-round-trip}, the OFSS functions during the round-trip protocol are straightforwardly obtained as ${\cal F}_M=2|c_1(\tau,\upsilon)|^2-1$ and {${\cal F}_A=|C_0(\tau,\upsilon)|$} respectively, although their analytical expression is cumbersome and therefore deferred to Appendix~\ref{app:OFSS-func}. The important outcome of their calculation is that they display a non-trivial dependence on $\tau_i$, in contrast with \eqref{eq:OFSS-func-M-single} and \eqref{eq:OFSS-func-A-single} obtained for a single passage (cf.~Figs.~\ref{fig:dependence-ti-M} and \ref{fig:dependence-ti-A}).\\

\begin{figure}[t]
\centering
		\includegraphics[width=8cm]{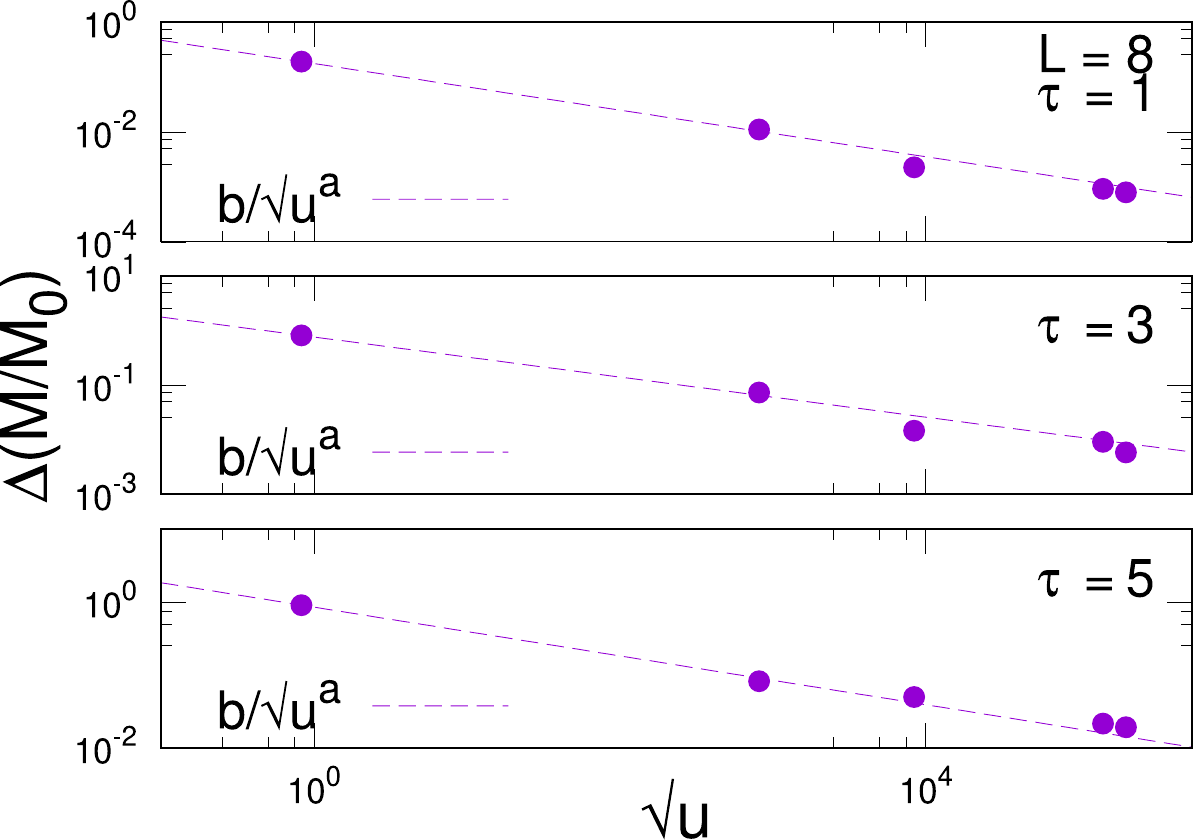}
\caption{Convergence to the OFSS of the longitudinal magnetization~---~Symbols show the quantity $\Delta(M/M_0)=|M/M_0-{\cal F}_M|$ for fixed $L=8$, $\upsilon=0.5$, $|\tau_i|=\tau_f=2$ (FOTs at $\tau=0,4$) and for different values of $\tau=1,3,5$ (different panels) as function of $\sqrt{u}$. Dashed line: power-law ansatz $f(u)=b/(\sqrt{u})^a$ with parameters $a$,$b$ extracted from a fit of the numerical data.}\label{fig:convergence-M}
	\end{figure}

\begin{figure}[t]
\centering
	\includegraphics[width=8cm]{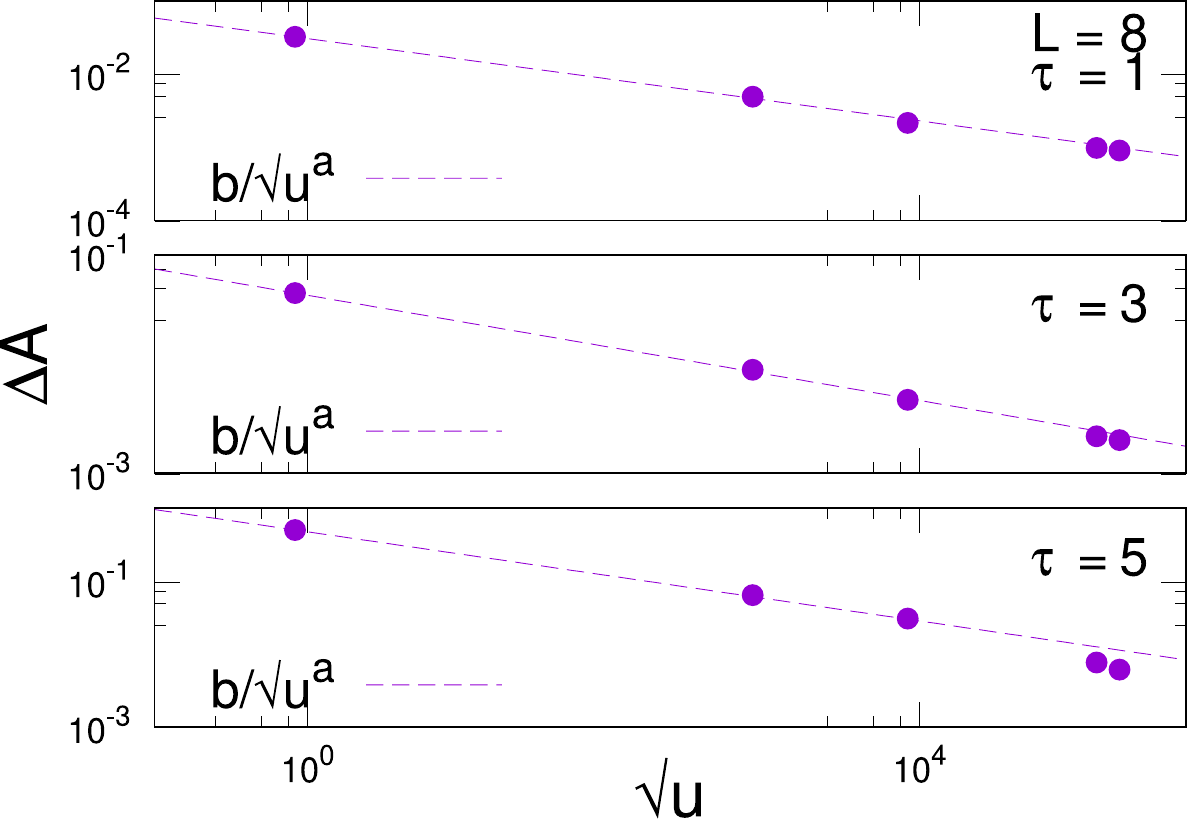}
		\caption{Convergence to the OFSS of the adiabaticity function~---~Symbols show the quantity $\Delta A=|A-{\cal F}_A|$ for fixed $L=8$, $\upsilon=0.5$, $|\tau_i|=\tau_f=2$ (FOTs at $\tau=0,4$) and for different values of $\tau=1,3,5$ (different panels) as function of $\sqrt{u}$. Dashed line: power-law ansatz $f(u)=b/(\sqrt{u})^a$ with parameters $a$,$b$ extracted from a fit of the numerical data.}
		\label{fig:convergence-A}
	\end{figure}

In Figs.~\ref{LongMagn} and \ref{AtripAt2u05g02}, the OFSS functions obtained from the solution of the LZS problem are plotted as dashed lines against the rescaled numerical data for the Ising model \eqref{eq:model}, 
showing an overall excellent agreement. Small deviations from the OFSS functions computed in the two-level approximation go monotonically to zero on increasing $t_{\rm KZ}=\sqrt{u}$, see Figs.~\ref{fig:convergence-M} and \ref{fig:convergence-A}.\\
 In these plots, our numerical analysis reveals a convergence to the OFSS behavior that is compatible with a power-law in $\sqrt{u}$. Further investigations on this aspect go beyond the scope of this work and thus are delivered to subsequent studies.

\subsection{Breakdown of the effective description}\label{sec:break}

	\begin{figure}[!h]
		\includegraphics[width=7cm]{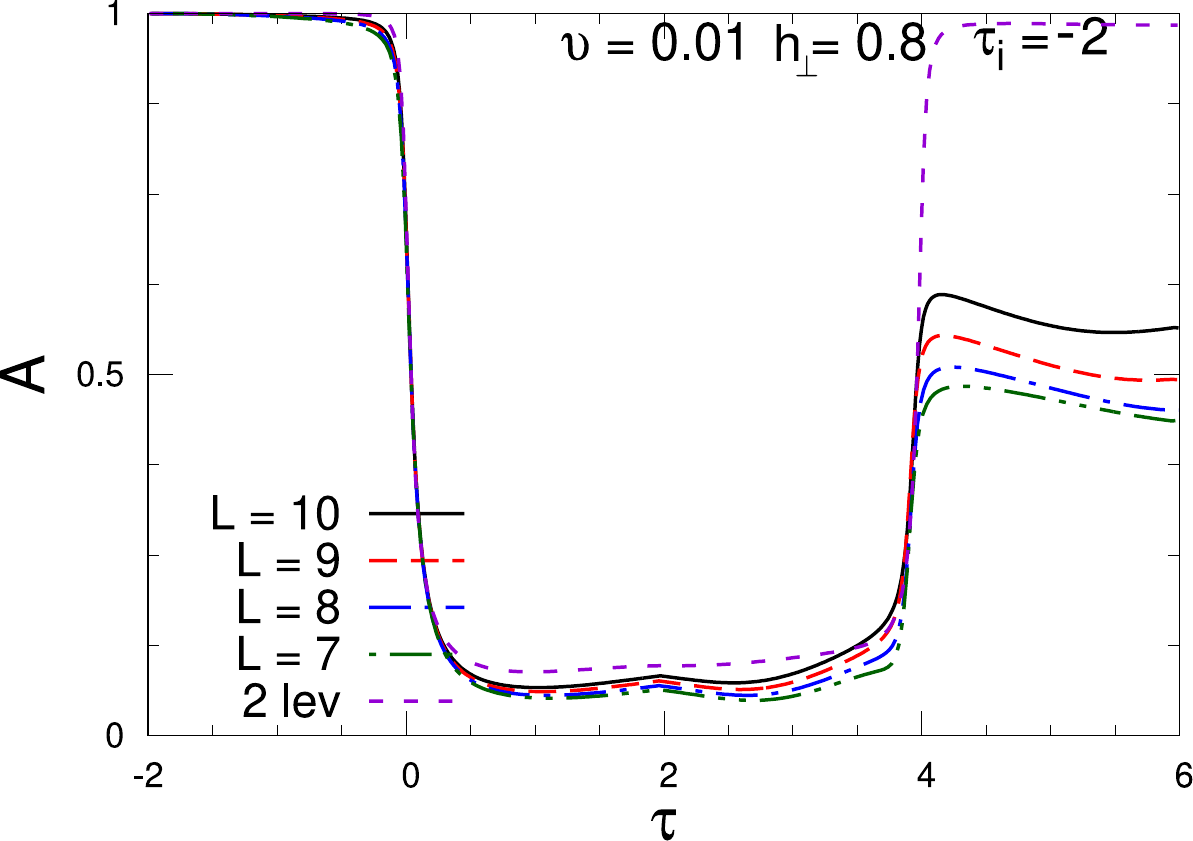}
		\caption{Adiabaticity function as a function of the rescaled time $\tau$ during a round-trip protocol with $|\tau_i|=\tau_f =2$ (FOTs at $\tau=0,4$). We set $\upsilon = 0.01$ and $h_\perp = 0.8$ and we vary the system size up to $L = 10$.  The dashed line shows the scaling function ${\cal F}_A(\tau,\upsilon)$ for the effective two-level model (see Sec.~\ref{sec:2-lev}).}
		\label{tripAt2u001g08}
	\end{figure}

	\begin{figure}[!h]
		\includegraphics[width=6cm]{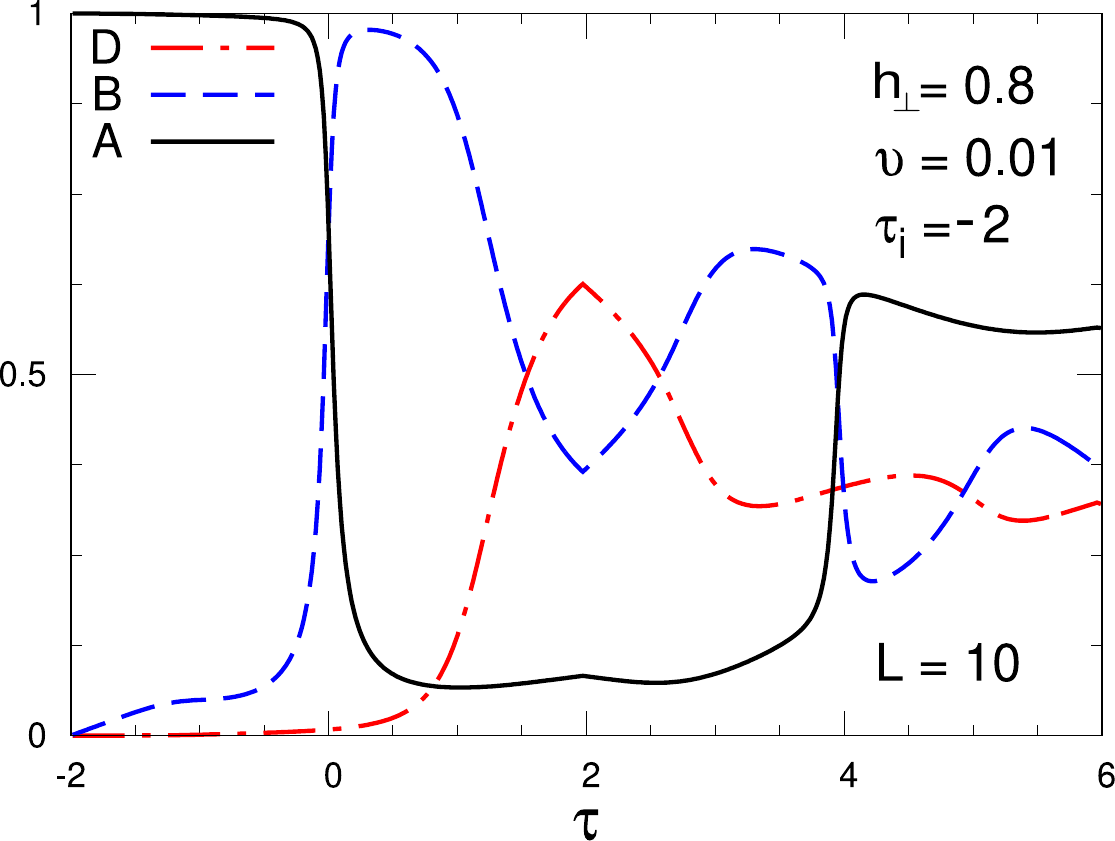}
		\caption{Behavior of $A(t)$, 
		$B(t)$ and $D(t)$ as function of the rescaled time $\tau$ 
		 for a fixed system size $L = 10$ during a round-trip protocol with $|\tau_i|=\tau_f=2$ (FOTs at $\tau=0,4$). We set $\upsilon = 0.01$ and $h_\perp = 0.8$.}
		\label{tripABDt2u001g08L10}
	\end{figure}
In this section, we show the breakdown of the OFSS regime (and consequently of the two-level effective description) discussed in Sec.~\ref{sec:OFSS-theory}. \\

As argued in Sec.~\ref{sec:effective-description}, the OFSS regime is able to capture the nonequilibrium dynamics arising by a slow driving across the quantum FOT for a time window where the dynamics involves the lowest two energy levels only. Corrections to the scaling behavior are expected when $\tau\sim{\cal O}(t_{\rm KZ})$. This is shown in Fig.~\ref{tripAt2u001g08} for the adiabaticity function and in Fig.~\ref{tripABDt2u001g08L10} for the distance measure. Here, $\upsilon=0.01$ and $t_{\rm KZ}\simeq 1.29$ for our choice of parameters, and we see that $D(\tau>t_{\rm KZ})\neq 0$, as expected. For other choice of parameters such that $\upsilon\sim{\cal O}(1)$ and $t_{\rm KZ}\gg 1$, we would observe the same qualitative behavior but occurring at larger time scales (cf.~ Eq.~\eqref{eq:energy-gap-01}).

\section{Floquet driving across a quantum FOT}\label{sec:Floquet}
As a natural extension of our setup, one can consider a Floquet driving of the Ising model \eqref{eq:model} across the quantum FOT, realized by a periodic repetition of the round-trip protocol discussed above. For sake of simplicity, we focus on the symmetric case $|t_i|=t_f\equiv t_0$  with time ramp
\be\label{eq:Floquet-ramp}
h_\parallel(t \ \text{mod} \ 4 t_0)=\begin{cases} t/t_s, \qquad t\in[-t_0,t_0]; \\[4pt] (2t_0-t)/t_s, \qquad t\in[t_0,3t_0].\end{cases}
\ee
Based on the arguments of Sec.~\ref{sec:break}, the validity of the two-level approximation is controlled by the energy injected in the model during the driving (and thus by the total duration of the protocol) rather than on the number of passages across the quantum FOT.\\

\begin{figure}[t]
\includegraphics[width=7cm]{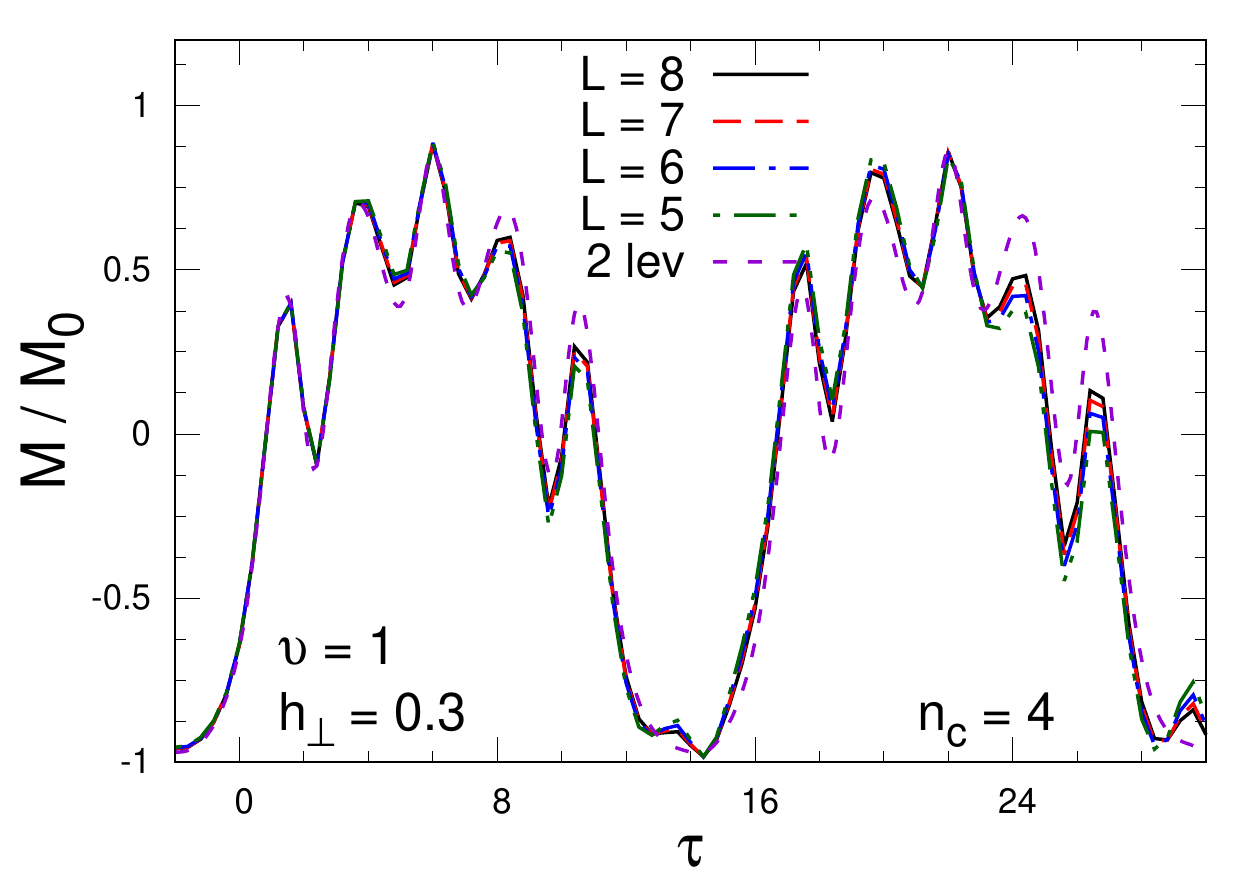}
\includegraphics[width=7cm]{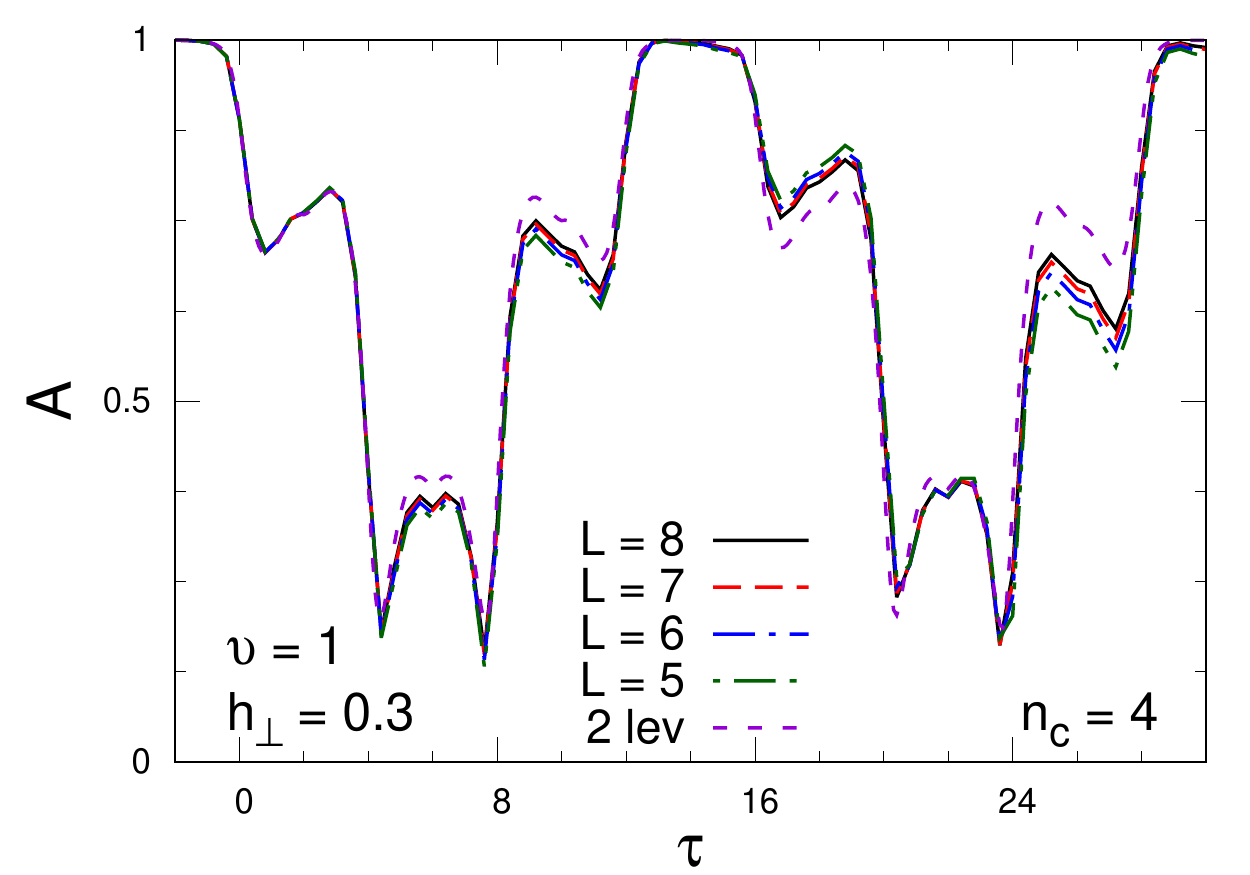}
\caption{OFSS during a Floquet driving of the Ising model \eqref{eq:model} across the quantum FOT (top~--~ $M/M_0$; bottom~--~$A$) as function of the rescaled time $\tau$. We set $\tau_0=2$ ($n_c=4$ round-trip cycles), $\upsilon=1$, $h_\perp=0.3$, and we vary the system size up to $L=8$. Dashed lines show the OFSS functions computed using the solution of the LZS problem \eqref{eq:LZS-round-trip}.}\label{fig:Floquet-M+A}
\end{figure}

 Therefore, we test the OFSS relations of Eqs.~\eqref{eq:OFSS-M}, \eqref{eq:OFSS-A} during the Floquet driving \eqref{eq:Floquet-ramp} of the Ising Hamiltonian \eqref{eq:model}. The results are shown in Fig.~\ref{fig:Floquet-M+A}, with OFSS functions computed by repeatedly applying the LZS solution \eqref{eq:LZS-round-trip} outlined in Sec.~\ref{sec:2-lev}. The figures show a good data collapse for different values of the system's parameters, with deviations from the OFSS functions (dashed lines) that increase with time. This behavior is compatible with the non-uniform convergence to the OFSS regime with time (i.e., larger values of $\tau$ require larger values of $u$), discussed in Sec.~\ref{sec:2-lev} and Sec.~\ref{sec:break}.\\

 We also notice that the system is driven extremely close to its initial state at time $\tau_\text{rec}\simeq 14$ for our choice of parameters (corresponding to a two periods driving).  Interestingly, this recurrence time is independent on the system size in terms of the OFSS variables. For $\tau>\tau_\text{rec}$, the pattern for the nonequilibrium dynamics of the quantities in Fig.~\ref{fig:Floquet-M+A} is repeated. This means that if we choose a sufficiently large (but finite) value of $t_{\rm KZ}$, the two-level effective model of Sec.~\ref{sec:2-lev} can provide a quite accurate description of the many-body system at any stage during the periodic driving. 
\\
\begin{figure}[t]
\centering
\includegraphics[width=6.5cm]{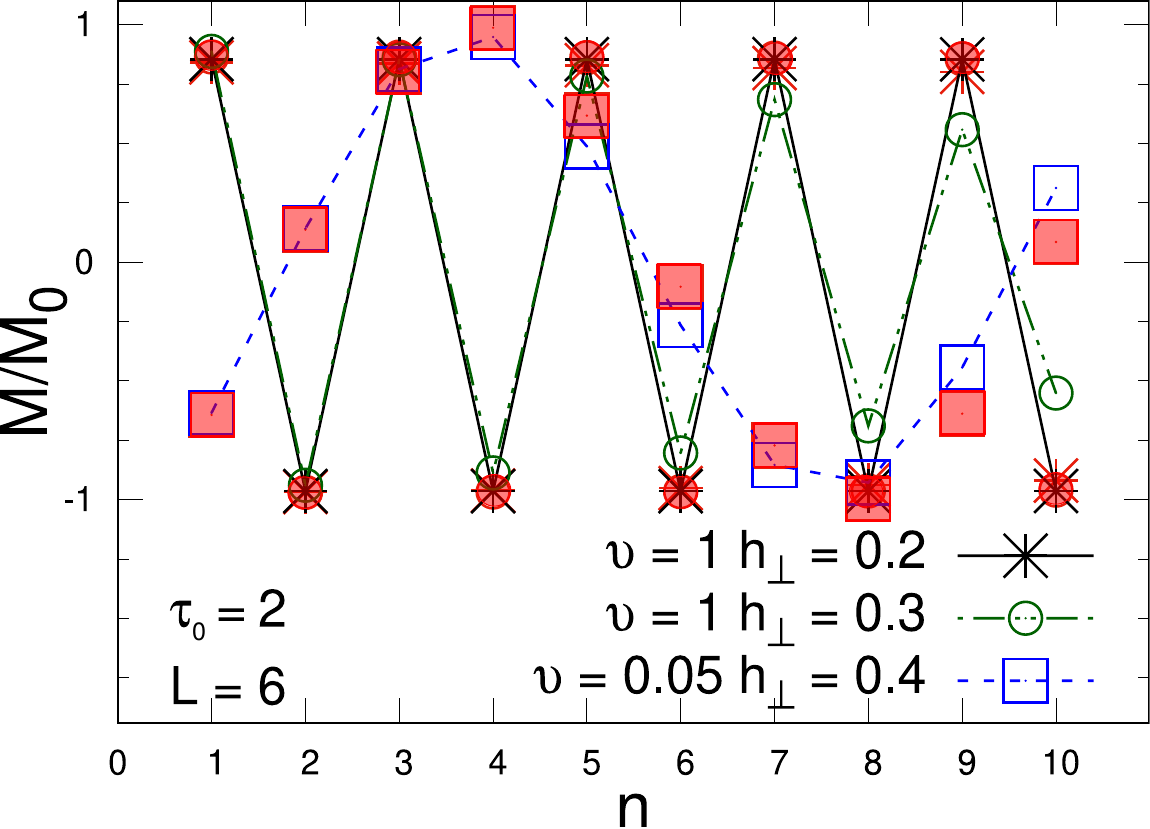}\\
\includegraphics[width=6.5cm]{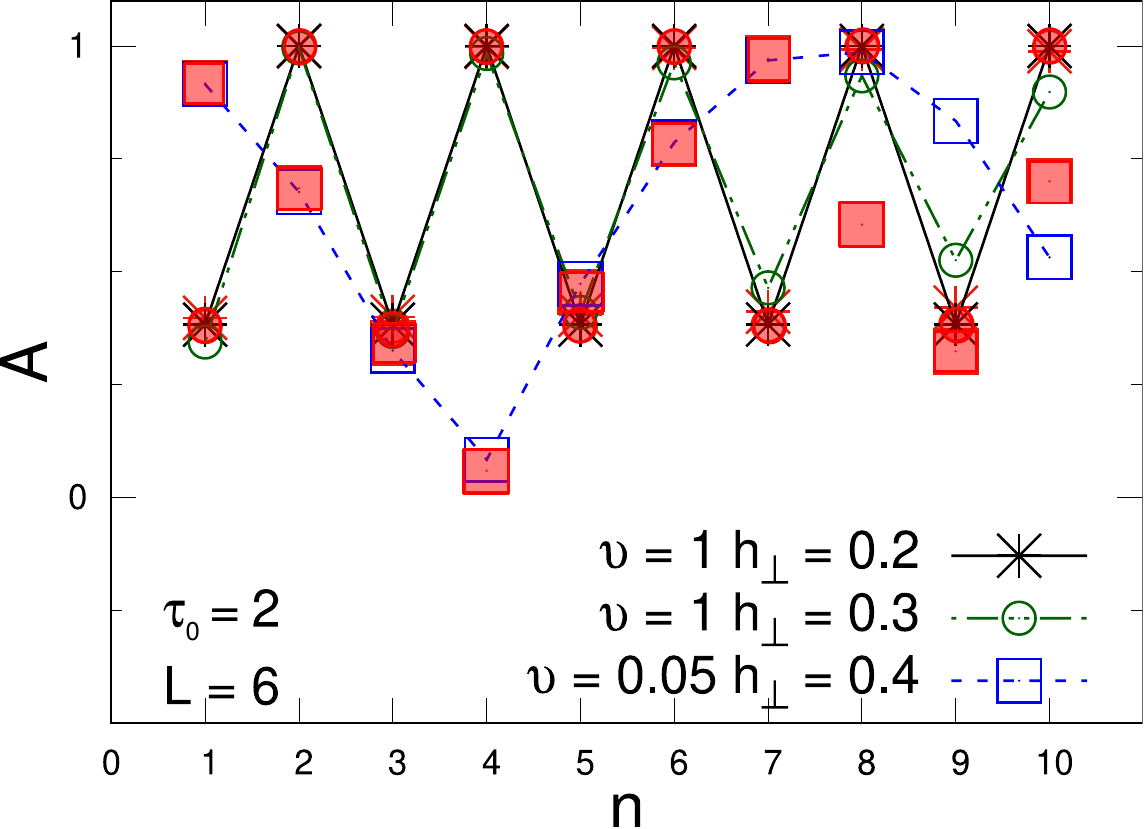}
\caption{Stroboscopic evolution of $M/M_0$ (top) and of $A$ (bottom) as function of  $n$ (corresponding to times $t=t_0(4n-1)$) at fixed $L=6$, $\tau_0=2$, and for {three different values of $\upsilon$, $h_\perp$ (different symbols)}. For each set of parameters, the filled symbols show the stroboscopic evolution in the two-level approximation.}\label{fig:strobo}
\end{figure}

\begin{figure}[t]\centering
\includegraphics[width=7cm]{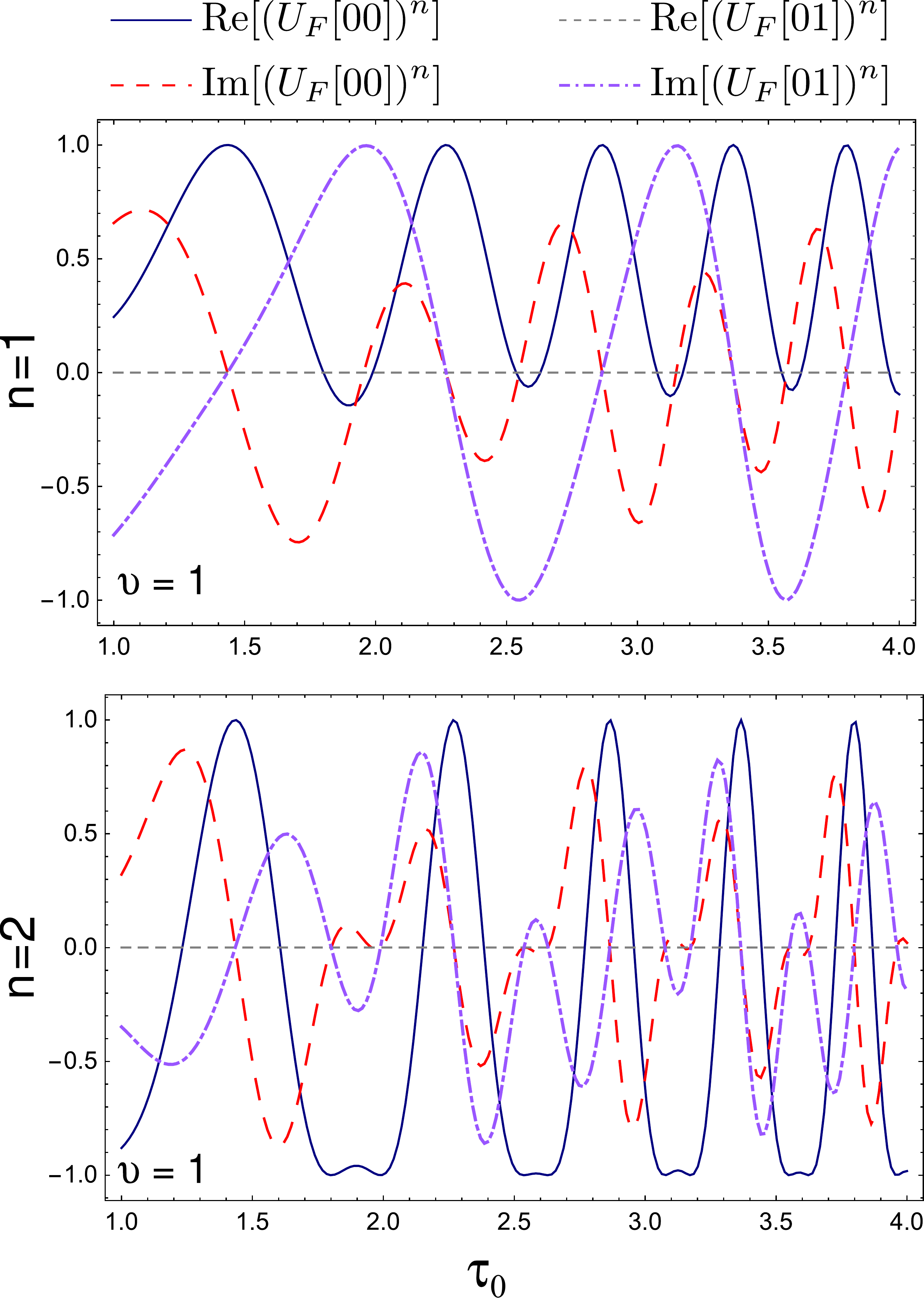}
\caption{{Matrix elements of $(U_F(\tau_0,\upsilon))^n$ as function of $\tau_0$, obtained from the analytical results of Appendix~\ref{app:OFSS-func}. In the panels, we set $\upsilon=1$ and $n=1$ (top), $n=2$ (bottom). }}\label{fig:rec}
\end{figure}
Finally, it is interesting to consider the evolution after $n$ periods, i.e., to look at the time-evolved wavefunction at stroboscopic times $t_n=t_0(4n-1)$. In the two-level approximation, this is given by
\be
\ket{\Psi(t_n)}\approx \left[(U_F)^n \ \begin{pmatrix} 1 \\[4pt] 0 \end{pmatrix}\right]^{\text{T}} \ \begin{pmatrix} \ket{\Psi_0[h_\parallel(t_0)]} \\[4pt] \ket{\Psi_1[h_\parallel(t_0)]} \end{pmatrix},
\ee
with Floquet evolution matrix $U_F=\tilde{U}(3\tau_0,\tau_0)\ U(\tau_0,-\tau_0)$. \\

In Fig.~\ref{fig:strobo}, we show the results for $M(t_n)$ and  $A(t_n)$, for a time window of $n=10$ periods. By comparing our numerical data with the two-level prediction (filled symbols), we observe an excellent agreement during the whole driving. In particular, we see that the convergence of our numerical data for the many-body system to the two-level prediction improves for fixed $\upsilon$ and $L$ on decreasing $h_\perp$ (i.e., on increasing the time scale $t_s$), as expected from the scaling arguments of Sec.~\ref{sec:OFSS-theory} and Sec.~\ref{sec:effective-description}. This is also confirmed by the curves obtained for $\upsilon=0.05$ and $h_\perp=0.4$ (square symbols), for which the value of $t_{\rm KZ}$ is much smaller. Accordingly, the curves initially show a good agreement with the OFSS theory but, for $n\gtrsim 8$, they significantly deviates from the OFSS functions. \\

{Notice also that the numerical analysis of Fig.~\ref{fig:strobo} clearly shows a two-period recurrence for both the longitudinal magnetization and the adiabaticity function. Such a peculiar behavior of the OFSS regime can be analytically investigated in the two-level approximation. Using the matrix elements of Appendix~\ref{app:OFSS-func}, it is easy to show that the equation
\be
(U_F(\tau_0,\upsilon))^n = \pm \mathds{1}
\ee
has solution for $n=2$, $\upsilon=1$ when $\tau_0= 2$, in agreement with what observed in Figs.~\ref{fig:Floquet-M+A} and \ref{fig:strobo}. In general, there exists a series of exceptional values of $\tau_0=\tau_\star(n,\upsilon)$ for which the system shows recurrence after $n$ periods, see Fig.~\ref{fig:rec}. For instance, $\tau_\star(1,1)\simeq 1.4$ for one round-trip protocol. Surprisingly, this means that the phases $\Phi(\tau_0,\upsilon)$ conferring to the system a non-trivial dependence on $\tau_0$ (see Appendix~\ref{app:OFSS-func}) can combine for some special values $\tau_0=\tau_\star(n,\upsilon)$ to restore the initial state of the system after $n$ periods. In this perspective, the aforementioned quasi-universality of OFSS regime can be exploited to engineer shortcuts to adiabaticity.
}

\section{Summary and conclusion}\label{sec:conclusion}
We investigate the unitary evolution of a $1D$ Ising model in a tilted magnetic field \eqref{eq:model}, in the presence of a time-dependent longitudinal field $h_\parallel=t/t_s$, $t_s\gg 1$ is the time scale, that drives the system through the quantum FOT in a round-trip fashion. We formulate the OFSS regime as the limit $L\to\infty$, $u\equiv t_s L^{-1} M_0^{-1}\to\infty$ where the time-dependent expectation values of local observables are proportional to quasi-universal OFSS functions of the variables $\tau=t/\sqrt{u}$ and $\upsilon=u \Delta^2(h_\perp, L)$. Here, the meaning of quasi-universality stands for the residual dependence of the OFSS functions on the details of the driving protocol at the inversion time. Numerical results for the many-body system confirm the validity of our scaling hypothesis (Figs.~\ref{LongMagn} and \ref{AtripAt2u05g02}). In Sec.~\ref{sec:effective-description}, we further probe the validity of the OFSS regime using time-dependent perturbation theory, relating it to the emergence of an effective two-level description which involves the lowest two states near the FOT. With this effective description, we reduce the driving protocol to a series of LZS transitions and we determine an analytical expression of the OFSS functions. Lastly, we extend the setup to the case of periodic driving across the quantum FOT, and we comment on the validity of the OFSS after several crossings. {Although our focus is on the Hamiltonian in Eq.~\eqref{eq:model}, we expect our results to apply to generic spin chains undergoing a quantum FOT, e.g., to quantum Potts chains \cite{campostrini2015finite} or to spin chains with staggering magnetic fields \cite{igloi2018quantum}.} \\

An interesting follow-up of this paper is the study of the round-trip protocol in the presence of weak dissipations, e.g. in the form of a  Lindblad dynamics for the quantum spin chain~\cite{rossini2020dynamic,rossini2019scaling,nigro2019competing,di2020dissipative,tarantelli2021quantum}. In this way, it might be possible to induce a relaxation of the system after each crossing and thus to investigate the OFSS of hysteresis cycles, similarly to what has been done, e.g., in Refs.~\cite{pelissetto2016off,scopa2018dynamical} for a thermal FOT under  relaxational dynamics.
\begin{acknowledgments}
SS acknowledges support from ERC under Consolidator grant No. 771536 (NEMO). SS is thankful to D.~Karevski for early-stage discussions that stimulate the development of this manuscript, and to P.~Fontana for collaboration on similar topics. The authors are grateful to E.~Vicari for useful discussions and critical remarks on the manuscript at various stages of its development.
\end{acknowledgments}
\appendix
\section{Equilibrium FSS functions}\label{app:eq-FSS-func}
In this appendix, we derive the FSS functions \eqref{eq:eq-scaling-func} for the Ising model \eqref{eq:model} at $h_\perp<1$, in the limit $h_\parallel\to 0^\pm$, $L\to\infty$. As noticed in Ref.~\cite{campostrini2014finite}, the quantum FOT is controlled by the competition of the two lowest energy levels. Therefore, it is possible to write down an effective two-level Hamiltonian by restricting the many-body system to the Hilbert space spanned by the ground state $\ket{\psi_0}$ and the first excited state $\ket{\psi_1}$, obtaining
\be
\hat{H}_\text{eff}=E_0 \hat{\mathds{1}}_{2\times 2} +\frac{1}{2}\left(\Delta(h_\perp,L) \hat\sigma^{(1)} -{\cal E}(h_\perp,h_\parallel,L)\hat\sigma^{(3)}\right),
\ee
where $E_0$ is the ground-state energy of the degenerate vacua at $h_\parallel=0$, $L=\infty$. This Hamiltonian is readily diagonalized in the basis
\begin{align}
&\ket{+}=\sin(\frac{\alpha}{2}) \ket{\psi_0} +\cos(\frac{\alpha}{2})\ket{\psi_1}\\
&\ket{-}=\cos(\frac{\alpha}{2}) \ket{\psi_0} -\sin(\frac{\alpha}{2})\ket{\psi_1}
\end{align}
with
\be
\tan(\alpha)=\kappa^{-1}=\frac{\Delta(h_\perp,L)}{{\cal E}(h_\perp,h_\parallel,L)}, \quad 0<\alpha\leq\frac{\pi}{2}.
\ee
It follows that the energy eigenvalues are given by
\be
E_\pm = E_0 \pm \frac{1}{2}\sqrt{{\cal E}^2+\Delta^2}
\ee
and, therefore, the energy gap \eqref{eq:eq-scalingE} reads as
\be
\Delta E(h_\perp,h_\parallel,L)=\Delta(h_\perp,L) \ \sqrt{1+\kappa^2}
\ee
from which one has $f_E(\kappa)=\sqrt{1+\kappa^2}$ (cf~Eq.~\eqref{eq:eq-scaling-func}). The longitudinal magnetization is 
\begin{align}
M(h_\perp,h_\parallel,L)&=M_0(h_\perp) \langle-|\hat\sigma^{(3)}|-\rangle
\nonumber \\[3pt]
&\quad =M_0(h_\perp) \left(\cos^2(\frac{\alpha}{2})-\sin^2(\frac{\alpha}{2})\right),
\end{align}
from which one finds the FSS function (cf.~Eq.~\eqref{eq:eq-scaling-func})
\be
f_M(\kappa)=\cos(\alpha)=\frac{\kappa}{\sqrt{1+\kappa^2}}.
\ee

\section{OFSS functions}\label{app:OFSS-func}
In this appendix, we focus on the solution of the finite-time LZS problem characterizing the linear driving of the Ising model \eqref{eq:model} across a quantum FOT, see Sec.~\ref{sec:2-lev}. For a better exposition, the case of single and of round-trip passages are treated in different subsections.
\subsection{Single passage}\label{app:LZS-single}
For a single passage through the quantum FOT starting at $\tau_i<0$, the finite-time LZS problem has an analytical solution~--~first derived in Ref.~\cite{vitanov1996landau}. The standard procedure is to decouple the set of equation \eqref{eq:LZS-single} by taking a time derivative. In this way, the differential equation, e.g.,~for $c_0(\tau,\upsilon)$ takes the form of a Weber differential equation
\be
\frac{d^2}{d\tau^2}c_0(\tau,\upsilon) + (\tau^2+\frac{\upsilon}{4}-i) c_0(\tau,\upsilon)=0,
\ee
which is solved in terms of Parabolic Cylinder functions. It follows that the $2\times 2$ evolution matrix $U(\tau,\tau_i)$ in \eqref{eq:LZS-single} has elements
\begin{widetext}
\begin{subequations}
\begin{align}
&U_{00}=\frac{\Gamma(1-\frac{i\upsilon}{8})}{\sqrt{2\pi}}\left[\mathscr{D}_{-1+\frac{i\upsilon}{8}}(\sqrt{2}e^{i\frac{3\pi}{4}}\tau_i)\mathscr{D}_{\frac{i\upsilon}{8}}(\sqrt{2}e^{-i\frac{\pi}{4}}\tau) + \mathscr{D}_{-1+\frac{i\upsilon}{8}}(\sqrt{2}e^{-i\frac{\pi}{4}}\tau_i) \mathscr{D}_{\frac{i\upsilon}{8}}(\sqrt{2}e^{i\frac{3\pi}{4}}\tau)\right];\\[4pt]
&U_{01}=\frac{2\Gamma(1-\frac{i\upsilon}{8})}{\sqrt{\pi\upsilon}}e^{i\frac{\pi}{4}}\left[\mathscr{D}_{\frac{i\upsilon}{8}}(\sqrt{2}e^{-i\frac{\pi}{4}}\tau_i)\mathscr{D}_{\frac{i\upsilon}{8}}(\sqrt{2}e^{i\frac{3\pi}{4}}\tau) - \mathscr{D}_{\frac{i\upsilon}{8}}(\sqrt{2}e^{i\frac{3\pi}{4}}\tau_i) \mathscr{D}_{\frac{i\upsilon}{8}}(\sqrt{2}e^{-i\frac{\pi}{4}}\tau)\right],
\end{align}
\end{subequations}
with $U_{10}=-U_{01}^*$ and $U_{11}=U_{00}^*$, $\Gamma(z)$ is the Euler Gamma function.\\
\end{widetext}
In the limit $|\tau_i|\gg 1$, these expressions can be simplified using known relations for $\mathscr{D}_\nu(z)$, reading as
\begin{subequations}\label{eq:U-large}
\begin{align}
&U_{00}(|\tau_i|\gg 1)\simeq e^{i\Phi(\tau_i,\upsilon)} e^{-\frac{\pi\upsilon}{32}} \mathscr{D}_{\frac{i\upsilon}{8}}(\sqrt{2}e^{i\frac{3\pi}{4}}\tau);\\[4pt]
&U_{10}(|\tau_i|\gg 1)\simeq e^{i\Phi(\tau_i,\upsilon)}\sqrt{\frac{\upsilon}{8}} e^{-i\frac{\pi}{4}} e^{-\frac{\pi\upsilon}{32}} \mathscr{D}_{-1+\frac{i\upsilon}{8}}(\sqrt{2}e^{i\frac{3\pi}{4}}\tau),
\end{align}
\end{subequations}
and the dependence on the initial condition $\tau_i$ drops out everywhere but in the phase
\be\label{eq:phase}
\Phi(q,\upsilon)=-q^2-\frac{\upsilon}{4}\log(\sqrt{2}|q|).
\ee
One can then easily obtain the expressions in Eqs.~\eqref{eq:OFSS-func-M-single} and \eqref{eq:OFSS-func-A-single} for the OFSS functions. The latter does not show any dependence on the initial condition $\tau_i$. It is interesting to note that by taking the limit $\tau\to\infty$ in Eq.~\eqref{eq:OFSS-func-M-single}, one obtains the Landau-Zener prediction for the defects abundance
\be
{\cal F}_M(\tau\to\infty,\upsilon)=1-2e^{-\frac{\pi\upsilon}{4}},
\ee
in agreement with standard Kibble-Zurek arguments \cite{dziarmaga2010dynamics}.
\subsection{Round trip}
With a similar strategy, it is possible to extend the solution of Appendix~\ref{app:LZS-single} to the round-trip protocol, by solving the two-level problem in the interval $t\in[t_f, 2t_f+|t_i|]$. Denoting with
\be
x=\frac{t_f+|t_i|-t}{\sqrt{u}},
\ee
one finds the following elements of the $2\times 2$ evolution matrix $\tilde{U}(\tau,\tau_f)$ in \eqref{eq:LZS-round-trip},
\begin{widetext}
\begin{subequations}
\begin{align}
&\tilde{U}_{00}=\frac{\Gamma(1-\frac{i\upsilon}{8})}{\sqrt{2\pi}}\left[\mathscr{D}_{\frac{i\upsilon}{8}}(\sqrt{2}e^{i\frac{3\pi}{4}}|\tau_i|)\mathscr{D}_{-1+\frac{i\upsilon}{8}}(\sqrt{2}e^{-i\frac{\pi}{4}}x) + \mathscr{D}_{\frac{i\upsilon}{8}}(\sqrt{2}e^{-i\frac{\pi}{4}}|\tau_i|) \mathscr{D}_{-1+\frac{i\upsilon}{8}}(\sqrt{2}e^{i\frac{3\pi}{4}}x)\right],\\[4pt]
&\tilde{U}_{10}=\frac{2\Gamma(1-\frac{i\upsilon}{8})}{\sqrt{\pi\upsilon}}e^{i\frac{\pi}{4}}\left[\mathscr{D}_{\frac{i\upsilon}{8}}(\sqrt{2}e^{i\frac{3\pi}{4}}|\tau_i|)\mathscr{D}_{\frac{i\upsilon}{8}}(\sqrt{2}e^{-i\frac{\pi}{4}}x) - \mathscr{D}_{\frac{i\upsilon}{8}}(\sqrt{2}e^{-i\frac{\pi}{4}}|\tau_i|) \mathscr{D}_{\frac{i\upsilon}{8}}(\sqrt{2}e^{i\frac{3\pi}{4}}x)\right],
\end{align}
\end{subequations}
with $\tilde{U}_{01}=-\tilde{U}_{10}^*$ and $\tilde{U}_{11}=\tilde{U}_{00}^*$. \\
\end{widetext}
Similarly to the previous case, these expressions simplify in the limit $|\tau_i|\gg 1$,
\begin{subequations}
\begin{align}
&\tilde{U}_{11}(|\tau_i|\gg 1)\simeq e^{i\Phi(\tau_i,\upsilon)} e^{-\frac{\pi\upsilon}{32}} \mathscr{D}_{\frac{i\upsilon}{8}}(\sqrt{2}e^{-i\frac{\pi}{4}}x),\\[4pt]
&\tilde{U}_{01}(|\tau_i|\gg 1)\simeq e^{i\Phi(\tau_i,\upsilon)}\sqrt{\frac{\upsilon}{8}} e^{-i\frac{\pi}{4}} e^{-\frac{\pi\upsilon}{32}} \mathscr{D}_{-1+\frac{i\upsilon}{8}}(\sqrt{2}e^{-i\frac{\pi}{4}}x).
\end{align}
\end{subequations}
From these equations (together with~\eqref{eq:U-large}), one finds
\begin{subequations}
\be\begin{split}\label{eq:coef0-round}
&c_0=e^{-\frac{\pi\upsilon}{16}}\Big\{\mathscr{D}_{\frac{i\upsilon}{8}}(\sqrt{2}e^{i\frac{3\pi}{4}}\tau_f)[\mathscr{D}_{\frac{i\upsilon}{8}}(\sqrt{2}e^{-i\frac{\pi}{4}}x)]^*\\[3pt]
&\qquad- \frac{i\upsilon}{8} e^{2i\Phi(\tau_i,\upsilon)}  \mathscr{D}_{-1+\frac{i\upsilon}{8}}(\sqrt{2}e^{i\frac{3\pi}{4}}\tau_f) \mathscr{D}_{-1+\frac{i\upsilon}{8}}(\sqrt{2}e^{-i\frac{\pi}{4}}x)\Big\},
\end{split}\ee
\be\begin{split}\label{eq:coef1-round}
&c_1=e^{-\frac{\pi\upsilon}{16}} \sqrt{\frac{\upsilon}{8}} e^{i\frac{\pi}{4}}\Big\{-\mathscr{D}_{\frac{i\upsilon}{8}}(\sqrt{2}e^{i\frac{3\pi}{4}}\tau_f)[\mathscr{D}_{-1+\frac{i\upsilon}{8}}(\sqrt{2}e^{-i\frac{\pi}{4}}x)]^*\\[3pt]
&\quad-i e^{2i\Phi(\tau_i,\upsilon)}  \mathscr{D}_{-1+\frac{i\upsilon}{8}}(\sqrt{2}e^{i\frac{3\pi}{4}}\tau_f) \mathscr{D}_{\frac{i\upsilon}{8}}(\sqrt{2}e^{-i\frac{\pi}{4}}x)\Big\},
\end{split}\ee
\end{subequations}
and determines the OFSS functions as detailed in the main text. \\
Notice that the coefficients \eqref{eq:coef0-round} and \eqref{eq:coef1-round} (hence the OFSS functions) keep a non-trivial dependence on the initial condition $\tau_i$ during the nonequilibrium dynamics via the phase \eqref{eq:phase}. This is not surprising given that $x_i\equiv |\tau_i|$ is the time at which the ramp is inverted after the first passage across the quantum FOT. At this time, the system is far from equilibrium and hence unable to wash out the memory on its initial (nonequilibrium) condition before being driven across the quantum FOT for the second time.

\bibliography{bibliography}

\end{document}